\documentclass{article}
\usepackage{amssymb}
\usepackage{epsfig}

\newcommand{\newc}{\newcommand}
\newc{\nf}{n_f}
\newc{\alphastar}{\alpha_s(q^*)}
\newc{\alphastartwo}{\alpha_s^2(q^*)}
\newc{\alphas}{\alpha_s}
\newc{\Vsub}{{\scriptscriptstyle V}}
\newc{\aVstar}{\alpha_\Vsub(q^*)}
\newc{\aVstarn}[1]{{\alpha_\Vsub^{#1}}(q^*)}
\newc{\aV}{\alpha_\Vsub}
\newc{\aVq}{\alpha_\Vsub(q)}
\newc{\aVmu}{\alpha_\Vsub(\mu)}
\newc{\aVmun}[1]{\alpha_\Vsub^{#1}(\mu)}
\newc{\logq}{\log(q^2)}
\newc{\logqtwo}{\log^2(q^2)}
\newc{\logqthree}{\log^3(q^2)}
\newc{\logaq}{\log((aq)^2)}
\newc{\logaqtwo}{\log^2((aq)^2)}
\newc{\logqn}[1]{\log^{#1}(q^2)}
\newc{\logqs}{\log(q^{*2})}
\newc{\logqsms}{\log(q^{*2}_\msbar)}
\newc{\logqsmsMb}{\log(q^{*2}_\msbar/\Mb^2)}
\newc{\logqsmsM}{\log(q^{*2}_\msbar/M^2)}
\newc{\logqstwo}{\log^2(q^{*2})}
\newc{\logqm}{\log(q^2/\mu^2)}
\newc{\logqqs}{\log(q^2/q^{*2})}
\newc{\logqqstwo}{\log^2(q^2/q^{*2})}
\newc{\logqqsthree}{\log^3(q^2/q^{*2})}
\newc{\logqsqn}{\log^n(q^{*2}/q^2)}
\newc{\logqmtwo}{\log^2(q^2/\mu^2)}
\newc{\logqsm}{\log(q^{*2}/\mu^2)}
\newc{\bzero}{\beta_0}
\newc{\bone}{\beta_1}
\newc{\btwo}{\beta_2}
\newc{\bthree}{\beta_3}
\newc{\msbar}{{\scriptscriptstyle {\rm \overline{MS}}}}
\newc{\MSbar}{{\overline{\rm MS}}}
\newc{\beq}{\begin{equation}}
\newc{\eeq}{\end{equation}}
\newc{\beqarray}{\begin{eqnarray}}
\newc{\eeqarray}{\end{eqnarray}}
\newc{\order}[1]{{{\cal O}(#1)}}
\newc{\dq}{d^4\!q\,}
\newc{\dnq}{d^n\!q\,}
\newc{\dnk}{d^n\!k\,}
\newc{\ave}{\int\dq}
\newc{\eq}[1]{Eq.~(\ref{#1})}
\newc{\eqs}[1]{Eqs.~(\ref{#1})}
\newc{\Ref}[1]{Ref.~\cite{#1}}
\newc{\Refs}[1]{Refs.~\cite{#1}}
\newc{\fig}[1]{Fig.~\ref{#1}}
\newc{\Table}[1]{Table~\ref{#1}}
\newc{\sect}[1]{Sect.~\ref{#1}}
\newc{\vev}[1]{\left\langle {#1}\right\rangle}
\newc{\Vev}[1]{\big\langle\!\!\big\langle {#1}\big\rangle\!\!\big\rangle}
\newc{\eps}{\epsilon}
\newc{\Idxdy}{\int_0^1\! dx\, dy\,} 
\newc{\Idx}{\int_0^1\! dx\,} 
\newc{\galpha}{{g^2\over (4\pi)^2}}
\newc{\half}{{1\over 2}}
\newc{\Mmsbar}{\overline{M}}
\newc{\Mb}{M_b}
\newc{\Mbmsbar}{\overline{M}_b}
\newc{\Mtau}{M_\tau}
\newc{\Mtaumsbar}{\overline{M}_\tau}
\newc{\Mt}{M_t}
\newc{\Mtmsbar}{\overline{M}_t}

\begin{document}

\title{
\hspace*{7.7cm} {\large hep-ph/0208224}\\
\vspace*{.6cm}
Scale Setting for $\alpha_s$ Beyond Leading Order
}

\author{
K.~Hornbostel$^a$, G.~P.~Lepage$^b$ and C.~Morningstar$^c$\\[.4cm]
\small $^a$Southern Methodist University, Dallas, TX 75275 \\
\small $^b$Newman Laboratory of Nuclear Studies,
            Cornell University, Ithaca, NY 14853 \\
\small $^c$Carnegie Mellon University, Pittsburgh, PA 15213  
}

\maketitle

\begin{abstract}
We present a general procedure for incorporating higher-order
information into the scale-setting prescription of Brodsky, Lepage 
and Mackenzie.  In particular, we show how to apply this prescription 
when the leading coefficient or coefficients in a series in the strong
coupling $\alphas$ are anomalously small and
the original prescription can give an unphysical scale.  
We give a general method for computing an optimum scale numerically, 
within dimensional regularization, and in cases when the coefficients 
of a series are known.  We apply it to the heavy quark mass and energy 
renormalization in lattice NRQCD, and to a variety of known series.  
Among the latter, we find significant corrections to the scales for the 
ratio of $e^+ e^-$ to hadrons over muons, the ratio 
of the quark pole to $\MSbar$ mass, the semi-leptonic $B$-meson
decay width, and the top decay width.  Scales for
the latter two decay widths, expressed in terms of $\MSbar$ masses,
increase by factors of five and thirteen, respectively, substantially
reducing the size of radiative corrections.
\end{abstract}

\pagebreak

\section{Introduction}

QCD processes computed to a finite order in perturbation theory
depend on both the choice of renormalization scheme and the scale 
for the running coupling constant $\alphas(q)$.  In particular, changes 
in the scale induce variations at the first neglected order.  While 
these variations diminish as higher orders are included, for low-order
calculations they can be significant, particularly for processes sensitive 
to relatively low scales.  Finding an optimum, physically motivated method 
for choosing this scale in such cases is important not only to produce
accurate results, but also to reasonably estimate convergence based 
on the size of the series terms.  Such a method allows a
meaningful prediction or comparison with data even at leading order.

A variety of procedures have been proposed to select this 
scale~\cite{Stevenson:1981du}--\cite{Brodsky:1997vq}.  
In this paper, we investigate the prescription 
of Brodsky, Lepage and Mackenzie (BLM)~\cite{Brodsky:1983gc}.  
In this method, one chooses the scale $q^*$ for $\alphastar$ 
which approximates the use of the fully dressed gluon propagator 
within that process.  The choice is equivalent to determining the 
dominant momentum flowing through the propagator within a 
diagram~\cite{Lepage:1993xa,Neubert:1995vb}.  It has been applied 
successfully in a large variety of perturbative calculations.
Among these, it was essential in demonstrating the viability of lattice
perturbation theory~\cite{Lepage:1993xa}, and in extracting a precise value 
of $\alphas$ from lattice simulations of the $\Upsilon$ and $\psi$ 
systems~\cite{El-Khadra:1994ss,Davies:1995ei}.

In this paper, we generalize the prescription to remedy an anomaly observed 
in a variety of applications, particularly apparent when determining the 
scale over a range of parameters in the action.  The NRQCD mass and energy 
renormalizations presented in \sect{sec:nrqcd} are typical examples.  In most 
of these cases, for some value of the bare quark mass, the BLM scale diverges.
We show that this breakdown is not a flaw in the general prescription, but 
rather the result of employing only a single vacuum-polarization insertion 
to estimate the typical momentum.  While we focus on setting the scale for 
one-loop diagrams, we use information from two-loop and higher insertions 
within these diagrams to provide a simple generalization which accurately 
estimates the scale over the full range of parameters.  It is straightforward 
to implement for both analytic and numerical computations, 
requiring only a modest extension beyond the leading order determination.
For both computations, one obtains the additional information required from 
one higher moment in $\log(q^2)$ within the same diagram as was used in the 
lowest order application.  For processes where the series coefficients are 
known, it requires only identifying the coefficient from vacuum polarization 
at the next order.  

We note that other authors have developed a variety of extensions to 
\Ref{Brodsky:1983gc}, which explore conformal symmetry and the relation 
between various perturbative 
schemes~\cite{Lu:1993nt,Brodsky:1995eh,Brodsky:2001cr}, or which estimate 
nonperturbative contributions and resum classes of diagrams to all 
orders~\cite{Beneke:1995qe,Neubert:1995vb}.  Our goal is more modest: to 
provide a simple but robust scale determination for a process calculated to 
finite order.  Specifically, we choose a single optimized scale for the 
leading, one-loop diagram, to be used for all orders.
We show, however, that our prescription should effectively absorb into
the leading term or terms the bulk of contributions from all higher order 
diagrams which dress the leading gluon.

\section{General prescription}
\label{sec:general}

\begin{figure}
\vspace{.2in}
\begin{center}
\epsfig{file=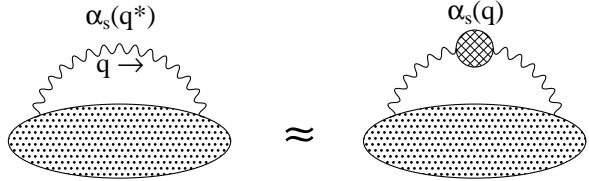, height=1.1in}
\end{center}
\caption{ 
  The BLM prescription for fixing the optimum scale $q^*$ 
  to leading order in $\alphastar$. }
\label{fd1}
\end{figure}

Following \Refs{Brodsky:1983gc,Lepage:1993xa}, we choose the $V$ scheme
based on the static-quark potential because of the direct connection 
between the scale of its coupling $\aV$ and the momentum flowing through its
associated gluon.
For a one-loop diagram with an integrand $f(q)$ which contributes
predominantly at large $q$, a natural choice 
for the scale $q^*$ of $\aV$ is a mean value which
reproduces the result of a fully dressed gluon within
the diagram~\cite{Brodsky:1983gc,Lepage:1993xa},
\beq
\label{main}
\aVstar \ave f(q) = \ave\,\aVq\,f(q)\; ,
\eeq 
as illustrated in \fig{fd1}.  
However, $\aVq$ possesses a pole at $\Lambda_V$, an artifact
of an all-orders summation of perturbative logarithms.  We avoid this
singularity by truncating the series for $\aVq$ at a finite order, 
as is appropriate for an asymptotic series.

Expanding $\aVq$ in terms of $\aVstar$
\beq
 \label{expand}
 \aVq = {\aVstar\over 1 + \aVstar\bzero \logqqs} \sim 
 \aVstar - \aVstarn{2} \bzero \logqqs + \cdots
\eeq
and solving to first nontrivial order gives~\cite{Lepage:1993xa}, 
\beq
\label{firstorder}
\logqs = {\ave f(q) \logq \over \ave f(q)} \equiv 
       {\vev{f(q)\logq}\over\vev{f(q)}} \equiv \Vev{\logq} \; ,
\eeq
a statement of this prescription suited for numerical calculations.
Here
\beq
\label{eq:b0def}
\bzero \equiv {1\over 4\pi}\left({11\over 3}\, C_A - 
{4\over 3}\, T_F\,\nf\right)
 = {1\over 4\pi}\left(11 - {2\over 3}\,\nf\right)\, , 
\eeq
and $\Vev{}$ indicates an average weighted by $f(q)$.

\begin{figure}
\vspace{.2in}
\begin{center}
\epsfig{file=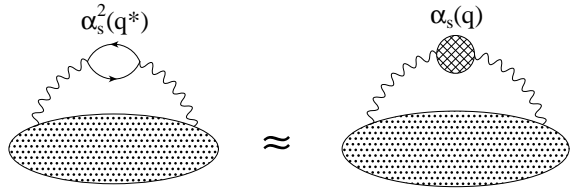, height=1.1in}
\end{center}
\caption{ 
  The BLM prescription applied to a process for which a gluon
  contributes first at order $\alphastartwo$. 
  The insertion on the left side represents vacuum polarization 
  from both quarks and gluons.}
\label{fd2}
\end{figure}

By the definition of $\aV$, \eq{firstorder} guarantees that $\aV(q^*)$ 
absorbs the effect of second-order vacuum polarization insertions in 
the gluon's propagator.  An alternate method to determine $q^*$ is then 
to require that $\bzero$, or equivalently $\nf$, disappears to that 
order~\cite{Brodsky:1983gc}.
This version is useful when the $\bzero$ or $\nf$ dependence of
coefficients in a perturbative expansion are known explicitly.
We discuss this in more detail in \sect{knownsrs}.

\eq{firstorder} produces an optimum scale by means 
of an average of $\log(q^2)$ weighted by $f(q)$.  As such, it provides
a measure of the typical momentum carried by this gluon in the
dominant integration region, in accord with intuition.
However, in certain cases $\vev{f}$ vanishes,
rendering $q^*$ from \eq{firstorder} meaningless.
This is a consequence of using an expression first-order in $\aVstar$
for a process which is properly second-order, rather than a flaw in the
general prescription.  When $\vev{f}$ vanishes, the diagram on the left
in \fig{fd1} does not contribute.  The leading contribution from
this gluon is second order, and the requirement that $q^*$ be chosen
to best approximate the all-order result leads to the equation illustrated 
in \fig{fd2}.  For an integrand dominated by large momentum, 
the left side of \eq{main} is replaced by
\beq
\label{eq:mainadd}
 -\aVstarn{2}\bzero \ave f(q)\logqqs \; ,
\eeq 
as is known from the running of $\aVq$.
Expanding $\aVq$ as in \eq{expand} yields
\beq
\label{secondorder}
\logqs =  {\vev{f\logqtwo}\over 2 \vev{f\logq}} \; .
\eeq
This, rather than \eq{firstorder}, is the appropriate statement of the 
prescription for this case.

As a simple illustration, consider a model in one dimension in which the 
Feynman diagram produces an integrand
\beq
\label{fdelta}
 f(q) = \delta(q - q_a) - \delta(q - q_b)\; ,
\eeq
with positive $q_a$ and $q_b$.  A reasonable expectation would be that
$q^*$ should be some average of the contributing scales
$q_a$ and $q_b$, particularly if they are nearby.  
Because $\vev{f}$ vanishes identically, \eq{firstorder} produces a 
divergent $q^*$, whereas \eq{secondorder}, which begins with the next
order contribution, gives for $q^*$ the more reasonable geometric mean
\beq
\label{fdeltaqstar}
 q^* = \sqrt{q_a q_b} \; .
\eeq
This is the same scale obtained by \eq{firstorder} applied to the positive
integrand
\beq
\label{fdeltasum}
 f(q) = \delta(q - q_a) + \delta(q - q_b)\; ,
\eeq
as might be expected.

In other cases, while not strictly vanishing, $\vev{f}$ may be 
anomalously small, and the dominant contribution from this gluon is still 
as in \fig{fd2}.  It is useful to generalize the lowest order prescription 
of \eq{firstorder} to incorporate both these cases naturally, and also to 
anticipate the situation where $\vev{f\log(q^2)}$ is anomalously small.  

The discrepancy between the left and right sides of \eq{main}
relative to the lowest order term $\aVstar \int\dq f(q)$ is
\beq
\label{errorone}
 - \aVstar\bzero\Vev{\logqqs} + \aVstarn{2}\bzero^2 \Vev{\logqqstwo}
\eeq
Applying \eq{firstorder} leaves a leading difference of
\beq
\label{erroronenext}
 \aVstarn{2}\bzero^2 \Vev{\left[\logq - \Vev{\logq}\right]^2}
  \equiv \aVstarn{2}\bzero^2 \sigma^2
\eeq
with $\sigma$ the standard deviation of $\logq$ with respect to 
the weight $f(q)/\vev{f}$.  When $f(q)$ does not 
change sign, $f(q)/\vev{f}$ and $\sigma^2$ are both positive.
When $f(q)$ changes sign and $\vev{f}$ is anomalously small
due to cancellations, this error can become
arbitrarily large, indicating that treating this as a first order process
is invalid and it is useful to incorporate information from the next order.

\begin{figure}
\vspace{.2in}
\begin{center}
\epsfig{file=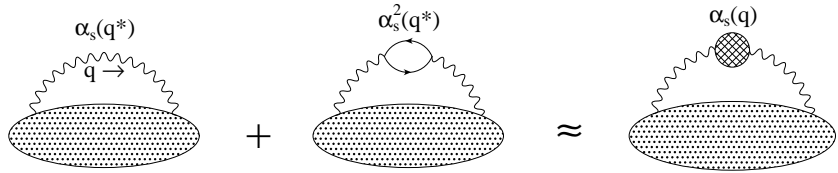, height=1.0in}
\end{center}
\caption{ 
  The BLM prescription applied to second order in $\alphastar$.
  The insertion on the left side represents vacuum polarization 
  from both quarks and gluons.}
\label{fd3}
\end{figure}

Matching the gluon's contribution to next order to the fully dressed
gluon, as in \fig{fd3}, adds the term in \eq{eq:mainadd} to the left side 
of \eq{main}.  Expanding both sides in terms of $\aVstar$ as before 
leads to a leading relative difference of 
\beqarray
\label{errortwo}
\aVstarn{2} \bzero^2 \Vev{\logqqstwo} = 
   \hspace*{6cm}\\[.1cm]
 -\aVstarn{2} \bzero^2 \left[\logqstwo - 2 \Vev{\logq}\logqs
 + \Vev{\logqtwo} \right]\; . \nonumber
\eeqarray
When $f(q)/\vev{f}$ is positive for all $q$, this discrepancy is also 
strictly positive, and the best that can be done is to choose $q^*$ 
to minimize it.  The result of minimization is again just
\eq{firstorder}.  This will also clearly be the case when $f(q)$ changes
sign in some small region without significant cancellations.  
The leading error is then the same as in \eq{erroronenext}.

However, when $f(q)$ possesses significant sign changes, the error in 
\eq{errortwo} can become negative for certain values of $\logqs$, 
and minimization is not appropriate.  In this case, it is possible to 
eliminate the difference altogether by choosing one of the two solutions
\beqarray
\label{qstarext}
 \logqs 
    &=& {\vev{f\logq} \pm \left[\vev{f\logq}^2 -
    \vev{f}\vev{f\logqtwo} \right]^{1\over 2} \over \vev{f}}\nonumber \\
   &\equiv& \Vev{\logq} \pm \left[\Vev{\logq}^2 - 
      \Vev{\logqtwo} \right]^{1\over 2} \nonumber \\
   &\equiv& \Vev{\logq} \pm \left[ - \sigma^2 \right]^{1\over 2} \, .
\eeqarray
When the logarithmic moments are available over a range of parameters, 
requiring that $q^*$ be continuous and physically sensible makes the 
proper choice apparent, as will be observed below.  In particular, 
when $\vev{f}$ is nearly zero, this requires choosing the sign opposite
to that of $\Vev{\logq}$.
In every case we have considered, the choice has been obvious.  
However, if the need arises, one may resolve the sign unambiguously by using 
information from higher moments, as discussed in \sect{highord}.  

Were $f(q)$ a probability distribution, $\sigma$ would be its standard
deviation.  A negative value for $\sigma^2$ indicates that $f(q)/\vev{f}$
has substantial changes of sign and is behaving significantly 
unlike a probability distribution.  In this case, $\vev{f}$ will be 
anomalously small, the order $\aVstarn{2}$ contribution becomes important, 
and \eq{qstarext} provides the appropriate choice of scale.
As a result, \eq{qstarext} determines $q^*$ when $\sigma^2$ is negative,
\eq{firstorder} when positive.  This prescription is the main result 
of this paper.

Although \eq{qstarext} uses information from part of the next
order, it is the appropriate scale to use when $\sigma^2$ is negative,
even if only computing to first order in $\aVstar$.  In that case, in 
addition to setting the scale for the leading term, it allows for a reasonable
estimate of the magnitude of the neglected next-order terms, based on
$\aVstarn{2}$.  When $\vev{f}$ is very small, these neglected terms
should give a sizable correction to the first-order term.  And when one
computes to order $\aVstarn{2}$ or higher using this scale, higher-order
terms which dress the leading gluon should be small, having been
largely absorbed into the first two, as in \fig{fd3}.

\begin{figure}
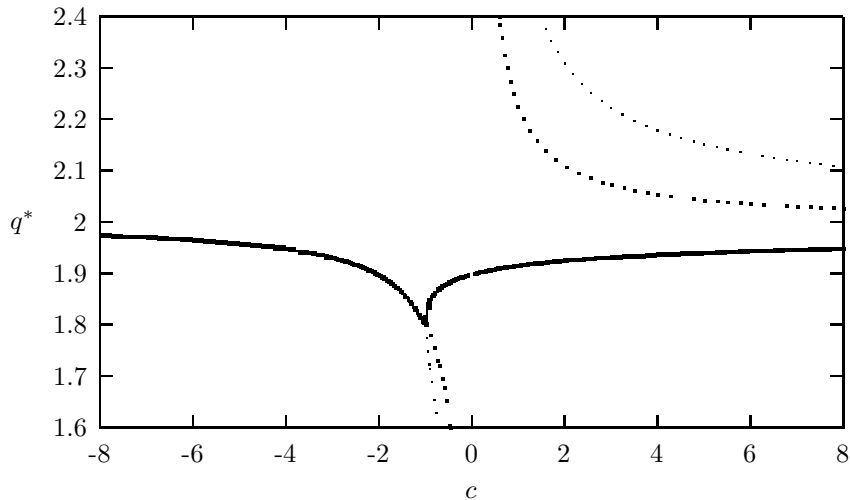

\include{plot0}
\caption{
The BLM scale $q^*$ for the model of \eq{fdelta2} as a function of $c$, 
with $q_a=2.0$ and $q_b=1.8$.  The first order solution of \eq{firstorder}
determines $q^*$ for $c<-1$, the second order solution of \eq{qstarext}
for $c>-1$.  The dark dotted lines show the first-order solution in 
regions in which it does not apply; light dotted lines display inapplicable 
second-order solutions.}
\label{deltaplot}
\end{figure}

As an illustration, we consider a slightly more 
general version of the model
of \eq{fdelta}, 
\beq
\label{fdelta2}
 f(q) = (1+c) \delta(q - q_a) - \delta(q - q_b)\; ,
\eeq
where $\vev{f}$ vanishes exactly for $c=0$, and has partial
cancellation for $c > -1$.  Figure~\ref{deltaplot} presents the scale 
$q^*$ determined by \eq{firstorder} when $\sigma^2 >0$ and by \eq{qstarext} 
when $\sigma^2 < 0$; that is, when $c > -1$.  For $c < -1$, with
no cancellations, \eq{firstorder} produces reasonable values
for $q^*$.  It falls between $q_a$ and $q_b$, approaching $q_a$ 
for $|c|$ large, and $q_b$ for $c=-1$.  
As $c$ approaches zero and $\vev{f}$ vanishes, $q^*$ from this
prescription diverges.  However, for $c > -1$, $\sigma^2$ is negative,
and \eq{qstarext} provides the optimum scale.  It evidently
behaves according to expectations.  Even for the case when $c$ is
large and positive and \eq{firstorder} produces a fairly sensible result,
it overestimates $q^*$ and \eq{qstarext} is preferable.

To summarize, we restate the prescription in a more compact 
form.  We have chosen inclusion of the running coupling within the 
integrand for a first-order diagram, $\ave\,\aVq\,f(q)$, as a natural means
to account for the running of the coupling with the gluon's momentum.  
It has the advantage that it incorporates higher-order diagrams which
dress the gluon, has no arbitrary scale dependence, and appropriately 
accounts for the strength of the coupling of a gluon with momentum $q$.

It has the disadvantage that $\aVq$ has an unphysical
pole at $q = \Lambda_V$, making the integrand ill-defined.  
We avoid this by expanding $\aVq$ at the scale $q^*$ as 
in \eq{expand}, and working to finite order in $\aVstar$, giving
\beq
\label{eqn:compact}
 \ave f(q)\left[\aVstar - \bzero\aVstarn{2}\logqqs 
   + \bzero^2\aVstarn{3}\logqqstwo + \cdots\right] \; .
\eeq
We choose $q^*$ to reproduce the full integral as well as possible.
In the absence of significant cancellations
in $\ave\, f(q)$, we may select the scale by \eq{firstorder} so that the first 
non-leading term in \eq{eqn:compact} vanishes.  The discrepancy 
is then the term of order $\aVstarn{3}$, which this choice for $q^*$ minimizes.
Furthermore, as $q^*$ will be near the typical $q$, $f(q)$ will be roughly 
even about $q^*$, and higher-order contributions should be either near
zero or their minimum depending on whether they are even or odd in $\aVstar$.

However, when $f(q)$ is essentially odd about some $q$ and so suffers
from significant cancellations, this is not an appropriate prescription.  
The leading term in \eq{eqn:compact} will be anomalously small compared
to the second term; in extreme cases, it might even vanish, and it would 
no longer make sense to absorb the second term into the leading term.
Furthermore, the scale from \eq{firstorder} would no longer accurately 
represent the typical momentum, and neglected higher order terms 
in \eq{eqn:compact} would be anomalously large.  It is, however, possible 
and sensible to require the third term to vanish by \eq{secondorder}; 
that is, to absorb it into the second.
This again provides a typical scale about which, in this case, $f(q)$ is 
essentially odd, minimizes the fourth-order term, and suppresses 
higher-order terms.

\section{Schemes other than $V$}
\label{sec:otherthanV}

For prescriptions other than $\aV$, vacuum polarization insertions
will in general contribute subleading constants in addition to
terms as in \eq{eq:mainadd}.  Though nonleading, these constants can make 
significant contributions at physically interesting values
of $q^2$, and so the optimum scale ought to be chosen to account for them as 
well.  One method for doing so is to focus on the fermion 
loop~\cite{Brodsky:1983gc}.  Both the $\log(q^2)$ and the subleading constant
will appear multiplied by $n_f$.  Replacing $n_f$ with $\bzero$ using
\eq{eq:b0def} modifies the fermion loop contribution by a constant, to 
\beq
-\alphastar\bzero(\logqqs + a)\; .
\eeq
When applying the first-order prescription, amending \eq{firstorder} to 
\beq
\label{shift}
\logqs = \Vev{\logq + a} \; 
\eeq
absorbs both the leading log and subleading constant into $\alphastar$.
For $\MSbar$, $a = -5/3$~\cite{Susskind:1976pi}--\cite{Buchmuller:1980bm}, 
resulting in the shift in scale~\cite{Brodsky:1983gc}
\beq
\label{eq:msVdiff}
 q_\msbar^* \; =\;  \exp(-5/6)\; q^*_\Vsub \; =\;  0.43\; q^*_\Vsub \, .
\eeq

As with $\aV$, this also absorbs the log associated with gluon vacuum 
polarization, since $\bzero$ determines its contribution relative to the 
fermion loop.  However, the gluonic subleading constant need not contribute 
in this ratio, and so will not also be completely absorbed.  One might choose 
instead to completely absorb the gluonic constant by solving \eq{eq:b0def}
for the adjoint Casimir constant $C_A = N$ associated with the gluon loop
in terms of $\bzero$, before absorbing the $\bzero$ term into $\alphastar$.
This would be particularly appropriate when $n_f=0$, for example.  For 
$\MSbar$, the result is a factor of $\exp(-31/66) = 0.63$, not greatly 
different from \eq{eq:msVdiff}.  This indicates that to one loop, absorbing 
the fermion loop constant also largely accounts for the gluonic constant.

When applying the second-order prescription, a constant subleading 
contribution leads to the same shift in \eq{qstarext} as in \eq{shift}, with
\beq
 \logqs = \Vev{\logq + a} \pm \left[\Vev{\logq}^2 - 
      \Vev{\logqtwo} \right]^{1\over 2}\; .
\eeq
The second term on the right is invariant under a shift in 
$\logq$ by a constant, and so remains unaffected.  

Because the $V$ scheme associates gluon exchange with a physical process at
the specific scale $q^{*2}$, higher order contributions associated with the
running of $\aV$ must vanish identically when the gluon's momentum
hits $q^{*2}$.  As a result, logarithmic contributions $\logqqs$
from these diagrams appear without subleading constants.

\section{Determining $q^*$ in $\MSbar$}

\begin{figure}
\vspace{.2in}
\begin{center}
\epsfig{file=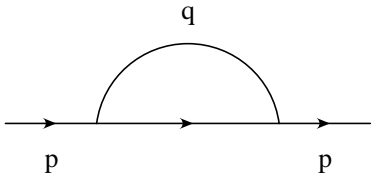, height=1.0in}
\end{center}
\caption{ 
 One loop self-energy diagram in the $g \phi^3$ model.}
\label{phi3}
\end{figure}

In order to provide a more realistic example, and to show how this 
prescription can be applied simply when using dimensional regularization,
we determine $q^*$ for the one-loop $g \phi^3$ diagram of \fig{phi3}.  
While we use this as a simplified model for a quark self-energy 
diagram, we also note that this scale setting method is not restricted to QCD.

By introducing an additional denominator of the form $(q^2)^\delta$ 
into this diagram in $n$-dimensional Euclidean space,
\beq
\label{vacpol}
 {g^2\over 2}\int {\dnq\over(2\pi)^n}\; {1\over q^2 + m^2}\;
 {1\over (p - q)^2 + m^2}\; {1\over (q^2)^\delta}\, ,
\eeq 
and expanding the
result in $\delta$ to second order, we produce the necessary 
logarithmic integrals:
\beq
\label{vacpolseries}
 \vev{f(q)} - \delta \vev{f(q) \log(q^2)} + 
   {\delta^2\over 2} \vev{f(q) \log^2(q^2)} + \, \ldots \, .
\eeq 
(Note that \Refs{Smith:1994id,Beneke:1995qe} present a different and elegant 
technique for extracting these logarithmic moments based on a simple 
dispersion sum over a fictional gluon mass.)

We evaluate \eq{vacpol} using standard methods and obtain
\beq
\label{vacpolDelta}
 \galpha {\Gamma(\delta + \eps)\over\Gamma(\delta)}
 \Idxdy y^{\delta - 1}(1-y) \left[4\pi\mu^2\over M^2\right]^\eps
 \left[1\over M^2\right]^\delta \, ,
\eeq
where $x$ and $y$ are the usual Feynman parameters, 
$\epsilon \equiv (4-n)/2$, $\mu$ is introduced
to keep $g$ dimensionless, and 
\beq
 M^2 \equiv (1-y) m^2 + x(1-y)(1-x+xy) p^2 \, .
\eeq

The integration region $y\sim 0$ produces a $1/\delta$ singularity.
Partial integration of $y^{\delta-1}$ makes this 
explicit, and allows us to expand in $\delta$ under the integral.  
Keeping terms in $1/\eps$ to order zero and comparing to \eq{vacpolseries}
gives
\beqarray
\label{delta0}
\lefteqn{\vev{f(q)} = -{1\over\eps} + 
   \Idx \log\left({m^2 + x(1-x) p^2\over\mu^2}\right)} \\[18pt]
\label{delta1} 
\lefteqn{\vev{f(q)\log(q^2)} = \Idxdy \bigg\{ -{1\over\eps^2} + 
   {1\over\eps}\log\left(y\over\mu^2\right) + 
   {\pi^2\over 12}  } \\[5pt]
   &&{} -\half\log^2\left(y\over\mu^2\right)
   + \half\log^2\left(M^2\over y\right) + 
     \left[1 - {x^2(1-y)^2 p^2\over M^2}\right] 
        \log\left(M^2\over y\right) \bigg\}  \nonumber \\[18pt]
\label{delta2} 
\lefteqn{\vev{f(q)\log^2(q^2)} = 2\Idxdy \bigg\{ -{1\over\eps^3} +
  {1\over\eps^2}\log\left(y\over\mu^2\right) } \\[5pt]
 &&{} + {1\over\eps}\left[{\pi^2\over 12} 
   - \half\log^2\left(y\over\mu^2\right)\right] 
  + {1\over 6}\bigg[-\psi^{\prime\prime}(1) - 
       {\pi^2\over 2}\log\left(y\over\mu^2\right) \nonumber \\[5pt]
 &&{} + \log^3\left(y\over\mu^2\right) + 
  3\left[1 - {x^2(1-y)^2 p^2\over M^2}\right]\log^2\left(M^2\over y\right)
    + \log^3\left(M^2\over y\right) \bigg] \bigg\}\, . \nonumber 
\eeqarray
In the above, we have dropped the overall factors $g^2/(4\pi)^2$, which are 
here irrelevant, and substituted $e^\gamma\mu^2/4\pi$ for $\mu^2$.  The latter 
greatly simplifies these expressions and allows us to apply the $\MSbar$ 
prescription by subtracting the $\eps$ poles alone, as one would for MS.

To renormalize these terms, we note that the $\log(q^2)$ and
$\log^2(q^2)$ factors integrated within this one-loop diagram 
stand in for the large-momentum contributions of the first two higher-order 
vacuum polarization subdiagrams which sum to form the running coupling.  In 
particular, as these factors are finite, they represent these subdiagrams
with their subdivergences already removed.  For example, the factor
$\log(q^2)$, which appears in the integrand which produced \eq{delta1}, 
comes from the large-momentum approximation to the $\MSbar$-renormalized 
one-loop vacuum polarization diagram; that is, to \eq{delta0} without the
pole.  The poles in $\eps$ which remain in Eqs.~(\ref{delta0}) 
to (\ref{delta2}) are then the new overall divergences associated
with one, two and three loops, respectively.  In the $\MSbar$ prescription 
these are simply discarded.

Finally, we note that \eq{delta0} in this model is the one-loop
vacuum polarization diagram in addition to being the particle's self-energy.
At large $p^2$, it is approximately $\log(p^2/\mu^2) - 2$, including
the subleading constant.  The constant $a$ in \eq{shift} is then $-2$, 
and the $\MSbar$ value for $q^*$ will differ by a factor $\exp(-1)$ from 
the expression for $q^*$ in the $V$ scheme.  

While \eqs{delta0} to (\ref{delta2}) allow us to determine $q^*$ for 
any $p$ and $m$, it is illuminating to consider two limits.
When $p^2 \gg m^2$, after renormalization and choosing $\mu^2 = p^2$
\beqarray
\label{largeP0}
\vev{f(q)} &=& -2\\[15pt]
\label{largeP1} 
\Vev{\log(q^2)} &=& \log(p^2) - {\pi^2\over 24} \\[15pt]
\label{largeP2} 
\Vev{\log^2(q^2)} &=& \log^2(p^2) - {\pi^2\over 12} \log(p^2)
 + \left[{\psi^{\prime\prime}(1)\over 6} - {\pi^2\over 12} + 2\right] 
\eeqarray
to leading order in $m^2/p^2$.  These give
\beq
\sigma^2 = 2 + {\psi^{\prime\prime}(1)\over 6} 
   - {\pi^2\over 12}\left[1 + {\pi^2\over 48} \right] \, ;
\eeq
or numerically, $\sigma^2=.6077$.  The lowest order solution is then 
appropriate and yields a value for $q^*$ close to $p$, with
\beq
 q^*/p = \exp(-1) \exp(-\pi^2/48) = .2995 \, ,
\eeq
which corresponds to a value of $\exp(-\pi^2/48) = .8141$ in the $V$ scheme.
We can use $\sigma$ to give a measure of the relative spread in the momenta 
which contribute to this diagram, with
\beq
 {\Delta q/q} \approx {\sigma/2} = .3898 \, ,
\eeq 
independent of the prescription.
In general, when $p^2\gg m^2$ but for $\mu^2$ arbitrary, 
$q^* \sim \sqrt{p \mu}$.  This is the same result as \eq{fdeltaqstar}, 
and reasonable for a diagram dominated by momenta between these two scales.

In the large mass limit $m \gg p$, with $\mu^2$ chosen to equal the 
natural scale $m^2$, we find
\beqarray
\label{largeM0}
\vev{f(q)} &=& {1\over 6}{p^2\over m^2}\\[15pt]
\label{largeM1} 
\vev{f(q)\log(q^2)} &=& \left({\pi^2-4\over 4}\right) 
       + {1\over 6}\log(m^2) {p^2\over m^2}\\[15pt]
\label{largeM2} 
\vev{f(q)\log^2(q^2)} &=& 
\left({\pi^2-4\over 4}\right) \left[1 +\log(m^2)\right]
    - {\psi^{\prime\prime}(1)\over 3}  \\[8pt]
    && {} + {1\over 3}\left[-1 + {\pi^2\over 6} 
       + {1\over 2}\log^2(m^2)\right] {p^2\over m^2} \nonumber
\eeqarray
to order $p^2/m^2$.  Clearly in this limit $\vev{f(q)}$ becomes anomalously 
small, and we expect the second order solution to be necessary.  We confirm 
this by noting that to this order
\beq
 \vev{f}^2 \sigma^2 = -\left({\pi^2-4\over 4}\right)^2
    + \left[ \left({\pi^2-4\over 12}\right)
   - {\psi^{\prime\prime}(1)\over 18}\right]{p^2\over m^2} 
\eeq
which is negative in this limit.  The second order formula applies, and gives
\beqarray
\label{MSsecond}
\log(q^{*2}/m^2) &=& -2 + \left[1 - {2\psi^{\prime\prime}(1)
\over 3(\pi^2-4)}\right] + \biggl[{\pi^2-3\over9(\pi^2-4)} \\[5pt]
  && {}  + {4\psi^{\prime\prime}(1)\over 27(\pi^2-4)^3}\bigl(-3(\pi^2-4) 
     + \psi^{\prime\prime}(1) \bigr)\biggr] {p^2\over m^2}\, , \nonumber 
\eeqarray
or numerically,
\beq
q^*/m = 0.6953 + 0.0574\ p^2/m^2 \, .
\eeq
For the $V$ scheme, the leading $-2$ in \eq{MSsecond} is absent, and 
\beq
q_\Vsub^*/m = 1.8899 + 0.1562\ p^2/m^2 \, .
\eeq

\begin{figure}
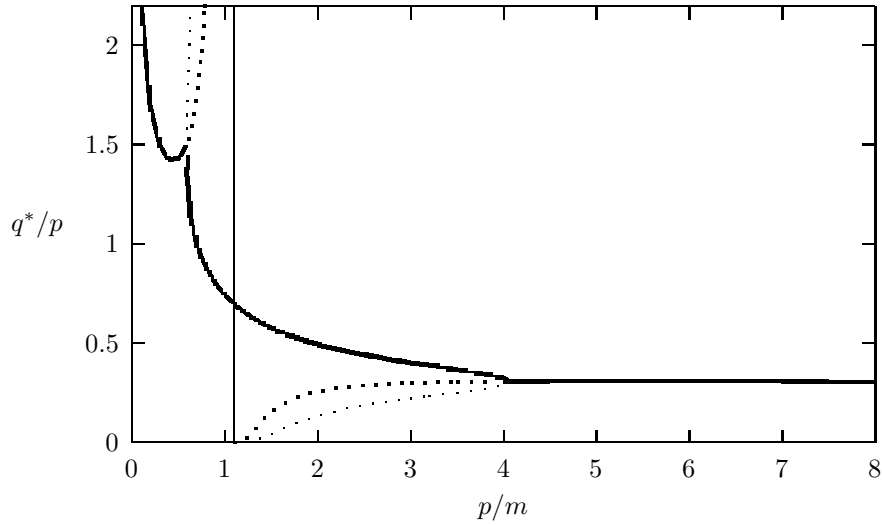

\include{plot3}
\caption{The $\MSbar$ BLM scale $q^*/p$ as a function of momentum $p/m$ 
for the diagram of \fig{phi3} in the scalar $\phi^3$ model, with  
$\mu = p$.  The first-order solution determines $q^*$ in both
the large and small $p$ regions, connected by the second-order solution
in the interim.  The vertical line indicates the point where the first-order 
solution diverges.  The dark dotted lines show the first-order solution in 
the region in which it does not apply; light dotted lines display inapplicable 
second-order solutions.
}
\label{plot3}
\end{figure}

\begin{figure}
\setlength{\unitlength}{0.240900pt}
\ifx\plotpoint\undefined\newsavebox{\plotpoint}\fi
\sbox{\plotpoint}{\rule[-0.200pt]{0.400pt}{0.400pt}}%
\begin{picture}(1349,809)(0,0)
\font\gnuplot=cmr10 at 10pt
\gnuplot
\sbox{\plotpoint}{\rule[-0.200pt]{0.400pt}{0.400pt}}%
\put(121.0,123.0){\rule[-0.200pt]{4.818pt}{0.400pt}}
\put(101,123){\makebox(0,0)[r]{0}}
\put(1308.0,123.0){\rule[-0.200pt]{4.818pt}{0.400pt}}
\put(121.0,255.0){\rule[-0.200pt]{4.818pt}{0.400pt}}
\put(101,255){\makebox(0,0)[r]{1}}
\put(1308.0,255.0){\rule[-0.200pt]{4.818pt}{0.400pt}}
\put(121.0,387.0){\rule[-0.200pt]{4.818pt}{0.400pt}}
\put(101,387){\makebox(0,0)[r]{2}}
\put(1308.0,387.0){\rule[-0.200pt]{4.818pt}{0.400pt}}
\put(121.0,519.0){\rule[-0.200pt]{4.818pt}{0.400pt}}
\put(101,519){\makebox(0,0)[r]{3}}
\put(1308.0,519.0){\rule[-0.200pt]{4.818pt}{0.400pt}}
\put(121.0,651.0){\rule[-0.200pt]{4.818pt}{0.400pt}}
\put(101,651){\makebox(0,0)[r]{4}}
\put(1308.0,651.0){\rule[-0.200pt]{4.818pt}{0.400pt}}
\put(121.0,783.0){\rule[-0.200pt]{4.818pt}{0.400pt}}
\put(101,783){\makebox(0,0)[r]{5}}
\put(1308.0,783.0){\rule[-0.200pt]{4.818pt}{0.400pt}}
\put(121.0,123.0){\rule[-0.200pt]{0.400pt}{4.818pt}}
\put(121,82){\makebox(0,0){0}}
\put(121.0,789.0){\rule[-0.200pt]{0.400pt}{4.818pt}}
\put(272.0,123.0){\rule[-0.200pt]{0.400pt}{4.818pt}}
\put(272,82){\makebox(0,0){1}}
\put(272.0,789.0){\rule[-0.200pt]{0.400pt}{4.818pt}}
\put(423.0,123.0){\rule[-0.200pt]{0.400pt}{4.818pt}}
\put(423,82){\makebox(0,0){2}}
\put(423.0,789.0){\rule[-0.200pt]{0.400pt}{4.818pt}}
\put(574.0,123.0){\rule[-0.200pt]{0.400pt}{4.818pt}}
\put(574,82){\makebox(0,0){3}}
\put(574.0,789.0){\rule[-0.200pt]{0.400pt}{4.818pt}}
\put(725.0,123.0){\rule[-0.200pt]{0.400pt}{4.818pt}}
\put(725,82){\makebox(0,0){4}}
\put(725.0,789.0){\rule[-0.200pt]{0.400pt}{4.818pt}}
\put(875.0,123.0){\rule[-0.200pt]{0.400pt}{4.818pt}}
\put(875,82){\makebox(0,0){5}}
\put(875.0,789.0){\rule[-0.200pt]{0.400pt}{4.818pt}}
\put(1026.0,123.0){\rule[-0.200pt]{0.400pt}{4.818pt}}
\put(1026,82){\makebox(0,0){6}}
\put(1026.0,789.0){\rule[-0.200pt]{0.400pt}{4.818pt}}
\put(1177.0,123.0){\rule[-0.200pt]{0.400pt}{4.818pt}}
\put(1177,82){\makebox(0,0){7}}
\put(1177.0,789.0){\rule[-0.200pt]{0.400pt}{4.818pt}}
\put(1328.0,123.0){\rule[-0.200pt]{0.400pt}{4.818pt}}
\put(1328,82){\makebox(0,0){8}}
\put(1328.0,789.0){\rule[-0.200pt]{0.400pt}{4.818pt}}
\put(121.0,123.0){\rule[-0.200pt]{290.766pt}{0.400pt}}
\put(1328.0,123.0){\rule[-0.200pt]{0.400pt}{165.257pt}}
\put(121.0,809.0){\rule[-0.200pt]{290.766pt}{0.400pt}}
\put(40,466){\makebox(0,0){$q^*/m\ \ $}}
\put(724,21){\makebox(0,0){$p/m$}}
\put(121.0,123.0){\rule[-0.200pt]{0.400pt}{165.257pt}}
\sbox{\plotpoint}{\rule[-0.500pt]{1.000pt}{1.000pt}}%
\multiput(342,809)(1.724,-20.684){3}{\usebox{\plotpoint}}
\multiput(347,749)(2.390,-20.617){4}{\usebox{\plotpoint}}
\multiput(355,680)(2.574,-20.595){2}{\usebox{\plotpoint}}
\multiput(362,624)(3.633,-20.435){3}{\usebox{\plotpoint}}
\multiput(370,579)(3.760,-20.412){2}{\usebox{\plotpoint}}
\put(382.03,521.51){\usebox{\plotpoint}}
\put(387.52,501.50){\usebox{\plotpoint}}
\multiput(393,483)(6.043,-19.856){2}{\usebox{\plotpoint}}
\put(407.35,442.55){\usebox{\plotpoint}}
\put(415.32,423.39){\usebox{\plotpoint}}
\multiput(423,409)(11.876,-17.022){3}{\usebox{\plotpoint}}
\multiput(453,366)(15.427,-13.885){2}{\usebox{\plotpoint}}
\put(495.84,330.44){\usebox{\plotpoint}}
\multiput(513,321)(19.044,-8.253){2}{\usebox{\plotpoint}}
\multiput(543,308)(20.097,-5.186){2}{\usebox{\plotpoint}}
\put(592.96,295.58){\usebox{\plotpoint}}
\multiput(604,293)(20.573,-2.743){2}{\usebox{\plotpoint}}
\put(654.48,286.27){\usebox{\plotpoint}}
\put(675.13,284.26){\usebox{\plotpoint}}
\multiput(694,283)(20.712,-1.336){2}{\usebox{\plotpoint}}
\put(725,281){\usebox{\plotpoint}}
\sbox{\plotpoint}{\rule[-0.600pt]{1.200pt}{1.200pt}}%
\put(725,281){\usebox{\plotpoint}}
\put(725,278.01){\rule{18.067pt}{1.200pt}}
\multiput(725.00,278.51)(37.500,-1.000){2}{\rule{9.034pt}{1.200pt}}
\put(875,278.01){\rule{18.308pt}{1.200pt}}
\multiput(875.00,277.51)(38.000,1.000){2}{\rule{9.154pt}{1.200pt}}
\put(951,279.51){\rule{18.067pt}{1.200pt}}
\multiput(951.00,278.51)(37.500,2.000){2}{\rule{9.034pt}{1.200pt}}
\put(1026,281.51){\rule{18.308pt}{1.200pt}}
\multiput(1026.00,280.51)(38.000,2.000){2}{\rule{9.154pt}{1.200pt}}
\put(1102,284.01){\rule{18.067pt}{1.200pt}}
\multiput(1102.00,282.51)(37.500,3.000){2}{\rule{9.034pt}{1.200pt}}
\put(1177,287.01){\rule{18.308pt}{1.200pt}}
\multiput(1177.00,285.51)(38.000,3.000){2}{\rule{9.154pt}{1.200pt}}
\put(1253,290.01){\rule{18.067pt}{1.200pt}}
\multiput(1253.00,288.51)(37.500,3.000){2}{\rule{9.034pt}{1.200pt}}
\put(800.0,280.0){\rule[-0.600pt]{18.067pt}{1.200pt}}
\sbox{\plotpoint}{\rule[-0.200pt]{0.400pt}{0.400pt}}%
\multiput(430,809)(2.865,-20.557){9}{\usebox{\plotpoint}}
\multiput(453,644)(5.034,-20.136){5}{\usebox{\plotpoint}}
\multiput(483,524)(7.288,-19.434){3}{\usebox{\plotpoint}}
\put(506.42,466.04){\usebox{\plotpoint}}
\multiput(513,452)(11.167,-17.495){3}{\usebox{\plotpoint}}
\multiput(543,405)(14.211,-15.128){2}{\usebox{\plotpoint}}
\multiput(574,372)(16.207,-12.966){2}{\usebox{\plotpoint}}
\multiput(604,348)(17.798,-10.679){2}{\usebox{\plotpoint}}
\put(649.55,321.74){\usebox{\plotpoint}}
\multiput(664,315)(18.808,-8.777){2}{\usebox{\plotpoint}}
\put(705.98,295.41){\usebox{\plotpoint}}
\put(717,288){\usebox{\plotpoint}}
\sbox{\plotpoint}{\rule[-0.600pt]{1.200pt}{1.200pt}}%
\put(129,215){\usebox{\plotpoint}}
\put(166,213.01){\rule{1.927pt}{1.200pt}}
\multiput(166.00,212.51)(4.000,1.000){2}{\rule{0.964pt}{1.200pt}}
\put(129.0,215.0){\rule[-0.600pt]{8.913pt}{1.200pt}}
\put(189,214.01){\rule{1.686pt}{1.200pt}}
\multiput(189.00,213.51)(3.500,1.000){2}{\rule{0.843pt}{1.200pt}}
\put(174.0,216.0){\rule[-0.600pt]{3.613pt}{1.200pt}}
\put(212,215.01){\rule{1.686pt}{1.200pt}}
\multiput(212.00,214.51)(3.500,1.000){2}{\rule{0.843pt}{1.200pt}}
\put(196.0,217.0){\rule[-0.600pt]{3.854pt}{1.200pt}}
\put(227,216.01){\rule{1.686pt}{1.200pt}}
\multiput(227.00,215.51)(3.500,1.000){2}{\rule{0.843pt}{1.200pt}}
\put(219.0,218.0){\rule[-0.600pt]{1.927pt}{1.200pt}}
\put(242,217.01){\rule{1.686pt}{1.200pt}}
\multiput(242.00,216.51)(3.500,1.000){2}{\rule{0.843pt}{1.200pt}}
\put(234.0,219.0){\rule[-0.600pt]{1.927pt}{1.200pt}}
\put(257,218.01){\rule{1.686pt}{1.200pt}}
\multiput(257.00,217.51)(3.500,1.000){2}{\rule{0.843pt}{1.200pt}}
\put(249.0,220.0){\rule[-0.600pt]{1.927pt}{1.200pt}}
\put(272,219.01){\rule{1.686pt}{1.200pt}}
\multiput(272.00,218.51)(3.500,1.000){2}{\rule{0.843pt}{1.200pt}}
\put(264.0,221.0){\rule[-0.600pt]{1.927pt}{1.200pt}}
\put(287,220.01){\rule{1.927pt}{1.200pt}}
\multiput(287.00,219.51)(4.000,1.000){2}{\rule{0.964pt}{1.200pt}}
\put(279.0,222.0){\rule[-0.600pt]{1.927pt}{1.200pt}}
\put(302,221.01){\rule{1.927pt}{1.200pt}}
\multiput(302.00,220.51)(4.000,1.000){2}{\rule{0.964pt}{1.200pt}}
\put(310,222.01){\rule{1.686pt}{1.200pt}}
\multiput(310.00,221.51)(3.500,1.000){2}{\rule{0.843pt}{1.200pt}}
\put(295.0,223.0){\rule[-0.600pt]{1.686pt}{1.200pt}}
\put(325,223.01){\rule{1.686pt}{1.200pt}}
\multiput(325.00,222.51)(3.500,1.000){2}{\rule{0.843pt}{1.200pt}}
\put(332,224.01){\rule{1.927pt}{1.200pt}}
\multiput(332.00,223.51)(4.000,1.000){2}{\rule{0.964pt}{1.200pt}}
\put(317.0,225.0){\rule[-0.600pt]{1.927pt}{1.200pt}}
\put(347,225.01){\rule{1.927pt}{1.200pt}}
\multiput(347.00,224.51)(4.000,1.000){2}{\rule{0.964pt}{1.200pt}}
\put(340.0,227.0){\rule[-0.600pt]{1.686pt}{1.200pt}}
\put(362,226.01){\rule{1.927pt}{1.200pt}}
\multiput(362.00,225.51)(4.000,1.000){2}{\rule{0.964pt}{1.200pt}}
\put(370,227.01){\rule{1.686pt}{1.200pt}}
\multiput(370.00,226.51)(3.500,1.000){2}{\rule{0.843pt}{1.200pt}}
\put(355.0,228.0){\rule[-0.600pt]{1.686pt}{1.200pt}}
\put(385,228.01){\rule{1.927pt}{1.200pt}}
\multiput(385.00,227.51)(4.000,1.000){2}{\rule{0.964pt}{1.200pt}}
\put(393,229.01){\rule{1.686pt}{1.200pt}}
\multiput(393.00,228.51)(3.500,1.000){2}{\rule{0.843pt}{1.200pt}}
\put(377.0,230.0){\rule[-0.600pt]{1.927pt}{1.200pt}}
\put(408,230.01){\rule{1.686pt}{1.200pt}}
\multiput(408.00,229.51)(3.500,1.000){2}{\rule{0.843pt}{1.200pt}}
\put(415,231.01){\rule{1.927pt}{1.200pt}}
\multiput(415.00,230.51)(4.000,1.000){2}{\rule{0.964pt}{1.200pt}}
\put(423,232.51){\rule{7.227pt}{1.200pt}}
\multiput(423.00,231.51)(15.000,2.000){2}{\rule{3.613pt}{1.200pt}}
\put(453,235.01){\rule{7.227pt}{1.200pt}}
\multiput(453.00,233.51)(15.000,3.000){2}{\rule{3.613pt}{1.200pt}}
\put(483,237.51){\rule{3.614pt}{1.200pt}}
\multiput(483.00,236.51)(7.500,2.000){2}{\rule{1.807pt}{1.200pt}}
\put(498,239.01){\rule{3.614pt}{1.200pt}}
\multiput(498.00,238.51)(7.500,1.000){2}{\rule{1.807pt}{1.200pt}}
\put(513,241.01){\rule{7.227pt}{1.200pt}}
\multiput(513.00,239.51)(15.000,3.000){2}{\rule{3.613pt}{1.200pt}}
\put(543,244.01){\rule{7.468pt}{1.200pt}}
\multiput(543.00,242.51)(15.500,3.000){2}{\rule{3.734pt}{1.200pt}}
\put(574,247.51){\rule{7.227pt}{1.200pt}}
\multiput(574.00,245.51)(15.000,4.000){2}{\rule{3.613pt}{1.200pt}}
\put(604,251.51){\rule{7.227pt}{1.200pt}}
\multiput(604.00,249.51)(15.000,4.000){2}{\rule{3.613pt}{1.200pt}}
\put(634,254.51){\rule{3.614pt}{1.200pt}}
\multiput(634.00,253.51)(7.500,2.000){2}{\rule{1.807pt}{1.200pt}}
\put(649,256.51){\rule{3.614pt}{1.200pt}}
\multiput(649.00,255.51)(7.500,2.000){2}{\rule{1.807pt}{1.200pt}}
\multiput(664.00,262.24)(2.469,0.505){4}{\rule{5.443pt}{0.122pt}}
\multiput(664.00,257.51)(18.703,7.000){2}{\rule{2.721pt}{1.200pt}}
\put(694,267.01){\rule{3.614pt}{1.200pt}}
\multiput(694.00,264.51)(7.500,5.000){2}{\rule{1.807pt}{1.200pt}}
\put(709,271.51){\rule{1.927pt}{1.200pt}}
\multiput(709.00,269.51)(4.000,4.000){2}{\rule{0.964pt}{1.200pt}}
\put(400.0,232.0){\rule[-0.600pt]{1.927pt}{1.200pt}}
\end{picture}
\caption{The $\MSbar$ BLM scale $q^*/m$ as a function of momentum $p/m$ 
for the diagram of \fig{phi3} in the scalar $\phi^3$ model, with  
$\mu = m$.  The first-order solution determines $q^*$ in 
the large $p$ region, while the negative root second-order solution gives 
$q^*$ for small $p$.  The dark dotted line shows the first-order solution in 
the region in which it does not apply; the light dotted line displays the
inapplicable second-order solution.
}
\label{plot4}
\end{figure}
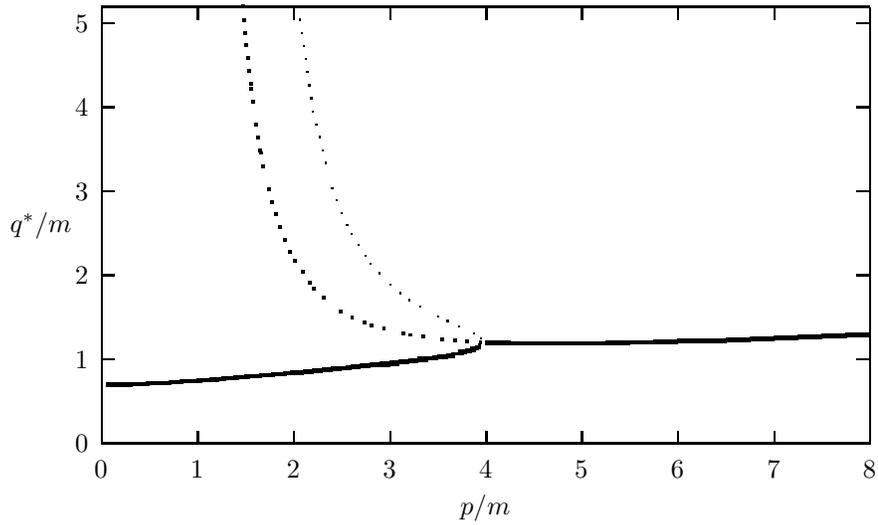

\fig{plot3} and \fig{plot4} display $q^*$ as a function of $p$ for the 
respective cases where $\mu = p$ and $\mu = m$.  The limiting values 
discussed above are evident.  For $\mu = p$, the first-order solution 
determines $q^*$ in both the large and small $p$ regions, connected by the 
second-order solution in the interim.  Immediately to the right of the point 
where the first-order solution diverges in \fig{plot3}, indicated by the 
vertical line, the second-order solution with positive root determines
$q^*$; to the left, the second-order negative root applies.  
For $\mu = m$ (\fig{plot4}),
the first-order solution applies for large $p$, the negative root 
second-order solution for small $p$.  In both cases, use of the
appropriate second order solution where applicable gives a meaningful
and continuous value for $q^*$ over the entire region in $p$.  Which
second-order solution to choose from \eq{qstarext} is obvious.

Finally, we note that computing higher order average logs for this diagram 
requires only expanding \eq{vacpolDelta} to higher orders in $\delta$, 
without the need to compute additional diagrams.

\section{Determining $q^*$ from a known series}
\label{knownsrs}

To apply this prescription we need the first two logarithmic moments 
within the integrand associated with a gluon's propagator.  Under certain 
conditions, we may apply this prescription to set the scale for a process
for which the expansion is already known by examining its 
$\nf$ dependence.  At each order
of $\aVmu$, the contribution from vacuum polarization will give 
the largest power of $\nf$, or equivalently, of 
$\bzero$~\cite{Beneke:1995qe,Neubert:1995vb,Brodsky:1995eh}.
It is therefore possible to read off 
the logarithmic integrals directly from the series coefficients.  

Using \eq{eq:b0def} to replace the largest-$n_f$ terms with $\bzero$ in 
contributions associated with a particular gluon, we obtain a series of 
the form
\beq
 c_0 \aVmu + (a_1 - c_1 \bzero) \aVmun{2} + 
   (a_2 + \cdots + c_2 \bzero^2) \aVmun{3} + \cdots \; .
\eeq
The coefficients $c_n$ are then associated with vacuum insertions in the
gluon propagator.
Comparison with the right sides of \eqs{main} and (\ref{expand}) gives
\beqarray
\label{cidents}
c_0 &=& \vev{f} \\
c_1/c_0 &\approx& \Vev{\logqm} \nonumber \\
c_2/c_0 &\approx& \Vev{\logqmtwo} \; ,\nonumber 
\eeqarray
which holds when $f(q)$ contributes predominantly at large $q$.
Given this association, the prescription to second order is
\beq
 \label{secondorderC}
 \logqsm = c_1/c_0  \pm \left[(c_1/c_0)^2 - c_2/c_0\right]^{1/2}
\eeq
when the argument of the square root is positive, and
\beq
 \label{firstorderC}
 \logqsm = c_1/c_0  
\eeq
otherwise.

For schemes other than $V$, the presence of a subleading constant $a$
contribution to fermion vacuum polarization leads to the identification
\beqarray
c_1/c_0 &\approx& \Vev{\logqm + a} \\
c_2/c_0 &\approx& \Vev{(\logqm + a)^2} \, .\nonumber 
\eeqarray
Because $c_1$ includes $a$ and the square root is insensitive
to it, \eq{firstorderC} and \eq{secondorderC} automatically incorporate
the shift in \eq{shift} and so may be used unchanged.  Also, if one is 
able to identify in the series the constants $C_A$ associated with gluonic
vacuum polarization, one could choose to use this instead to rewrite the
series in terms of $\bzero$.  \eq{firstorderC} and \eq{secondorderC} would
then automatically absorb the subleading gluonic constant,
as discussed at the end of \sect{sec:general}.

\section{Combining series}
\label{combining}

Determining the scale for the series formed by multiplying two series,
\beq
 F_a = 1 + c_a \aV(q^*_a) + \cdots
\eeq
and 
\beq
 F_b = 1 + c_b \aV(q^*_b) + \cdots
\eeq
with known scales is straightforward when considering only first-order scale
setting:
\beq
 F_{ab} \equiv F_a F_b = 1 + (c_a + c_b) \aV(q^*_{ab}) + \cdots
\eeq
with
\beq
\log(q^{*2}_{ab}) = {c_a \log(q^{*2}_a) + c_b \log(q^{*2}_b)
\over c_a + c_b} \, .
\eeq

Because of the need to first test the sign of the new $\sigma_{ab}^2$,
the prescription for applying second-order scale setting is slightly
more involved.  In that case,
\beqarray
\sigma_{ab}^2 &=& {c_a \Vev{\logqtwo}_a + c_b \Vev{\logqtwo}_b 
\over c_a + c_b} -
 \left({c_a \Vev{\logq}_a + c_b \Vev{\logq}_b 
\over c_a + c_b}\right)^2 \nonumber\\
 &=& {c_a \sigma^2_a + c_b \sigma^2_b \over c_a + c_b} +
 {c_a c_b\over (c_a + c_b)^2}\left(\Vev{\logq}_a - \Vev{\logq}_b\right)^2 \, .
\eeqarray
As usual, if $\sigma_{ab}^2 > 0$, the first order combination 
\beq
\log(q^{*2}_{ab}) = {c_a \Vev{\logq}_a + c_b \Vev{\logq}_b\over c_a + c_b} \, .
\eeq
applies.  If $\sigma_{ab}^2 < 0$, 
\beq
\log(q^{*2}_{ab}) = {c_a \Vev{\logq}_a + c_b \Vev{\logq}_b\over c_a + c_b}
   \pm \left[-\sigma_{ab}^2\right]^\half \, .
\eeq

When combining two series by division, $F_{a/b}\equiv F_a/F_b$, the first-order
coefficients subtract rather than add.  The above formulas again apply, but
with the replacement $c_b \rightarrow -c_b$.  These also apply to series
combined by addition and subtraction, respectively, because the results at
first order are equivalent.

For schemes other than $V$, one should amend the average logs to include
the subleading constants, as discussed in \sect{sec:otherthanV}.  The relation
between the scale in $V$ and in other schemes remains the same.

\section{Higher orders}
\label{highord}

Extending this prescription beyond second order is relatively 
straightforward, though it requires computation of $\Vev{\logqthree}$ 
and higher moments, or information from third-order and higher terms in 
a known series.  An extension to third order, for example, would be 
necessary should both $\vev{f}$ and $\vev{f\logq}$ vanish,
making the third term in \eq{eqn:compact} the leading term.  Absorbing 
the subsequent term by requiring $\vev{f\logqqsthree}$ to vanish would give
\beq
\label{eqn:torder}
\logqs =  {\vev{f\logqthree}\over 3 \vev{f\logqtwo}} \; , 
\eeq
as one would also obtain from \fig{fd2} with two
loops on the left side.

When $\vev{f}$ and $\vev{f\logq}$ are not identically zero but
are anomalously small relative to higher moments, requiring 
$\vev{f\logqqsthree}$ to vanish still gives the appropriate scale.
A symptom that this is the case would be a third-order scale
near \eq{eqn:torder} which shifts significantly the scale
obtained at lower order, and which is more in line with
physical expectations.  If available, higher even moments near
their minima or odd moments near zero at this scale would
confirm it.

An order-$n$ equation is necessary to set $q^*$ 
only when all of the first $(n-2)$ moments vanish or are anomalously
small.  It would be unusual for a generic integrand $f(q)$ to be 
effectively orthogonal to more than a few powers of $\logq$,
and so the need to use a high-order equation should be rare.
We have found no realistic cases for which either \eqs{firstorder} or 
(\ref{qstarext}) were not sufficient.

\begin{figure}
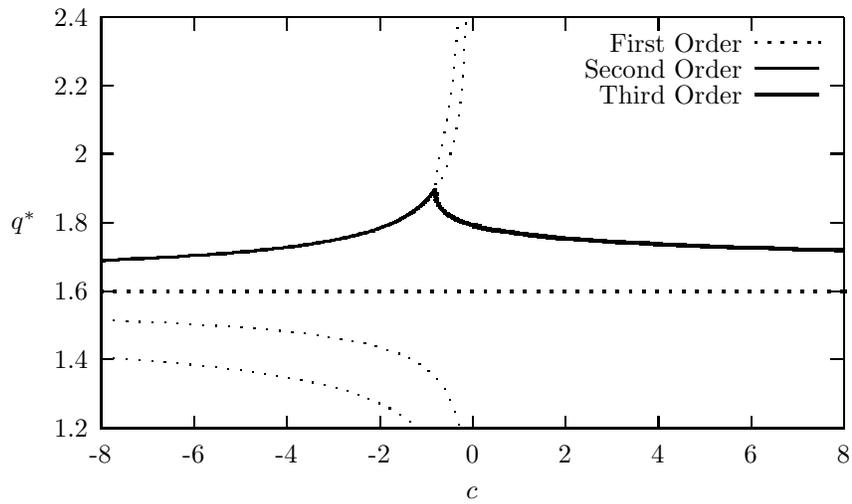

\include{cubic3}
\caption{
The BLM scale $q^*$ for the model of \eq{fdelta3} as a function of $c$, 
with $q_a=2.0$, $q_b=1.8$ and $q_d=1.6$.  The second-order solution 
determines $q^*$ for $c<-1$, the third-order solution
for $c>-1$.  The dark dotted line shows the first-order solution;
light dotted lines display inapplicable second and third-order solutions.
}
\label{cubic3}
\end{figure}

In \fig{cubic3} we illustrate the appropriate scales for a model,
\beq
\label{fdelta3}
 f(q) = A \delta(q - q_a) + B \delta(q - q_b) + (D + c) \delta(q - q_d)\; ,
\eeq
with
\beq
 B = -A \log(q_a/q_d)/\log(q_b/q_d), \;\;\; D = -(A + B)\; ,
\eeq
contrived such that both $\vev{f}$ and $\vev{f\logq}$ vanish at $c=0$.
The second- and third-order solutions behave as expected where appropriate;
the first-order solution, while not divergent, is low throughout.
The unphysical behavior of the first-order solution, as well as the
significant discrepancy between first- and third-order scales 
indicate that the first-order result is inadequate.

Note the one- and two-node structures of the integrand $f(q)$ in the
two- and three-delta models of \eq{fdelta2} and \eq{fdelta3}, 
similar to generic first and second excited-state wavefunctions.  
This is the result of choosing $f(q)$ to be orthogonal to the zeroth,
and to both the zeroth and first powers of $\logq$, respectively.  
Integrands requiring higher-order equations would necessarily have 
more nodes and additional detailed structure.

\begin{figure}
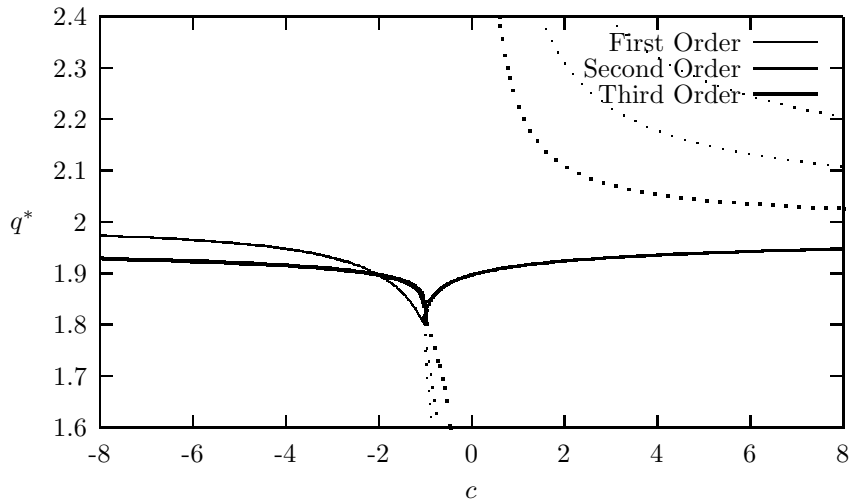

\include{cubic2}
\caption{
The BLM scale for the $\delta$-function model of \eq{fdelta2} as in
\fig{deltaplot}, but with the inclusion of the third-order solution.
It is applicable for $c<-1$ and indicated by the additional dark line.  
Dotted lines indicate inapplicable solutions.
}
\label{cubic2}
\end{figure}

In general, higher-order solutions can confirm that a scale determined
at a lower order is indeed typical.  For example,
in \fig{cubic2} we include the third order solution for the \eq{fdelta2}
model; where applicable, it does not differ significantly from the 
first-order result.  This will also be apparent when we examine higher 
moments for certain processes in \sect{applications}.  

Thus far, for simplicity, we have restricted our discussion to 
contributions to $\aVq$ from one loop vacuum polarization.  We show 
here that the above expressions for $q^*$ are not limited to these. 
Specifically, including subleading contributions to $\aVq$
expanded within a diagram as in \eq{eqn:compact} gives
\beqarray
\label{anseriesC}
&&\aVstar\vev{f} + \aVstarn{2}\big[\bzero\Delta_1\big] 
   + \aVstarn{3}\big[\bzero^2\Delta_2 + \bone\Delta_1\big] \\
&&\ \ \ \ \ \ \ \ \ \ 
   {}+ \aVstarn{4}\big[\bzero^3\Delta_3 + {5\over 2}\bzero\bone\Delta_2
            +\btwo\Delta_1\big] \nonumber\\
&&\ \ \ \ \ \ \ \ \ \ 
   {}+ \aVstarn{5}\big[\bzero^4\Delta_4 + {13\over 3}\bzero^2\bone\Delta_3
          + 3\bzero\btwo\Delta_2 + {3\over 2}\bone^2\Delta_2 
          + \bthree\Delta_1 \big] + \cdots \, . \nonumber
\eeqarray
Here
\beq
\bone \equiv {1\over (4\pi)^2}\left(102 - {38\over 3}\,\nf\right)\; , 
\eeq
and
\beq
\btwo \equiv {1\over (4\pi)^3}
             \left({2857\over 2} - {5033\over 18}\,\nf + 
               {325\over 58}\,\nf^2\right)\, ,
\eeq
and we have defined the moments
\beq
\label{deldef}
   \Delta_n \equiv \vev{f\logqsqn} 
    = \vev{f\left[\logqs - \logq\right]^n}  \, .
\eeq

We see from \eq{anseriesC} that the reasoning which led to the 
prescriptions in \eq{firstorder} and \eq{qstarext} was not dependent on 
the leading-log approximation to $\aVq$: the diagrams in each new set
introduced by increasing the order of the leading-log approximation
have as coefficients the same sequence of moments $\Delta_n$
as in the leading approximation.  For non-anomalous
cases, requiring $\Delta_1$ to vanish via \eq{firstorder} absorbs the
first leading-log correction, proportional to $\beta_0\Delta_1$, into the 
first term.  It also absorbs the first next-to-leading log contribution,
proportional to $\beta_1\Delta_1$, and so on for each logarithmic order.
And in each order, it leaves a leading discrepancy proportional to the 
moment $\Delta_2$, which is minimized by \eq{firstorder}.  Insofar as
this $q^*$ represents the typical momentum carried by this gluon,
higher-order terms proportional to higher moments should be suppressed.
As a result, the first term $\aVstar\vev{f}$, with the choice of a
single scale for $q^*$, should do reasonably well at approximating the 
right side of \eq{main} regardless of the number of loops kept in the
$\beta$ function for $\aVq$.  
When $\vev{f}$ vanishes or is anomalously small, it is inappropriate to 
absorb the leading correction at each logarithmic order into the
vanishing or small first term.  In this case \eq{qstarext} gives the 
appropriate scale.  It causes $\Delta_2$ to vanish, and so absorbs the 
second correction at each logarithmic order into the set of leading 
corrections.  Again, because \eq{qstarext} produces a typical scale, 
higher-order terms should be small.

This illustrates one of the advantages of using a coupling based
on a physical process, such as $\aV$.  For other schemes, there can
be significant subleading constants associated with the diagrams
which dress the gluon, not accounted for in the running coupling.
These will appear, for example, in the coefficients associated with 
the leading log term at each order $n$; that is, with $\bzero^{n-1}$, as
\beq
\label{deldefsub}
   \Delta_n \equiv \vev{f\left[\logqs - (\logq + a_0)\right]^n} \; ,
\eeq
and one may absorb them by adjusting the scale.
However, only one such constant can be absorbed in this manner; parts
of those associated with $\bone$ and higher will remain.
By its definition, these constants cannot appear for $\aV$, and $\aVstar$ 
well represents the strength of a physical gluon at
the scale $q^*$.  This optimum choice of scale minimizes all coefficients
associated with dressing the gluon, not just those associated with $\bzero$.

\section{Improving convergence}
\label{improving}

Thus far we have been considering the appropriate scale $q^*$
for the leading $\aV$.  For the sake of simplicity, we will be 
content to optimize the scale for the leading term, and will
use the same scale for higher-order diagrams.  From the previous 
discussion, it is clear that this will be a reasonable scale
for diagrams which dress the leading-order gluon;
these contributions should be small, having been largely absorbed
into the leading term or terms.
There is no reason to believe, however, that it will be the
best scale for other higher-order diagrams, and it should
certainly be possible to improve the convergence of the series
by choosing the scale for such diagrams 
separately~\cite{Brodsky:1983gc,Brodsky:1995eh}.

There are other cases for which it could prove advantageous to
allow different scales at different orders, for different
diagrams within the same order, or even within a single diagram.
Returning to the simple and unexceptional model of \eq{fdeltasum}, 
for which the first-order scale setting \eq{firstorder} gives 
\eq{fdeltaqstar}, we note that the moments
\beq
 \Vev{\logqsqn} = \half\left[\log^n(q^{*2}/q_a^2)
       +\log^n(q^{*2}/q_b^2) \right]
\eeq
become zero for $n$ odd, and $\log^n(q_a/q_b)$ for $n$ even.
The latter grow in magnitude when the $q_a$ is a few times greater
than $q_b$. As these are proportional to the coefficients of terms
which dress the gluon, terms in the series become $\approx 1$
when this range $q_a/q_b$ exceeds a few over $\alphas$.  The series 
becomes badly behaved then not only when the scale for $\alphas$ 
is low, but also when the range of important momenta is 
large~\cite{Beneke:1995qe,Neubert:1995vb}.

Nevertheless, in this simple case the remedy is obvious.  The problem
results from requiring $\alphas$ at a single scale to incorporate
two widely different scales.  Separating these and writing the series 
for this model in terms of both $\alphas(q_a)$ and $\alphas(q_b)$ incorporates 
vacuum polarization contributions exactly and causes higher moments 
to vanish.  This reinforces the idea that for a series in which 
gluons from different diagrams occur in
loops sensitive to significantly different momenta, allowing the $\alphas$
associated with each to have its own scale could improve the 
series' convergence.  Furthermore, for a series in which a gluon 
in a single diagram is 
sensitive to a wide range of momenta one might even consider improving its
behavior by splitting up the integrand, with $\alphas$ at a different scale 
assigned to each region, as in the example above.

Even for diagrams which dress a specific gluon, it is possible
to minimize higher moments by allowing the scale associated with 
these diagrams to differ at different orders.  
In particular, one may 
select the scale at every other order such that the following moment 
vanishes, and the next is minimized.  Variations in the scale would
account for the different regions the moments probe in the integrand.

We will not pursue this further here, but mention that care must be
taken to preserve gauge invariance if separate scales are assigned to
different parts of a series.

\section{Applications}
\label{applications}

\subsection{Known series}

In \Table{aptable} we present a collection of results for perturbative
quantities for which at least the second logarithmic moment is available, 
allowing us to apply scale setting beyond lowest order.  
\Ref{Brodsky:1995eh} presents a useful compilation and discussion 
of many of these.  These include
the $\log$ of the $1\times 1$ Wilson loop in lattice QCD ($-\log W_{11}$)
\cite{Lepage:1993xa,Klassen:1995mw},
the ratio of $e^+ e^-$ goes to hadrons over muons ($R_{e^+ e^-}$)
\cite{Chetyrkin:1979bj}--\cite{Surguladze:1991tg},
the ratio of the quark pole mass to its $\MSbar$ mass ($M/\overline{M}$)
\cite{Tarrach:1981up}--\cite{Melnikov:2000qh} and 
\cite{Beneke:1995qe,Neubert:1995vb},
the ratio of $\tau$ goes to $\nu_\tau$ + hadrons over
$\tau$ goes to $\nu_\tau e^- \overline{\nu_e}$ ($R_\tau$)
\cite{Braaten:1988hc}--\cite{Pivovarov:1992rh} and
\cite{Gorishnii:1991vf,Surguladze:1991tg,Ball:1995ni},
the semileptonic $B$-meson decay width 
($\Gamma(B \rightarrow X_u e \overline{\nu})$)
\cite{Berman:1958ti}--\cite{vanRitbergen:1999gs}
expressed both in terms of the pole and $\MSbar$ $b$-quark masses, 
the top quark decay width ($\Gamma(t \rightarrow b W )$)
\cite{Cabibbo:1979jg}--\cite{Czarnecki:1995jt} and 
\cite{Smith:1994id,Jezabek:1989iv,Beneke:1995qe}
in terms of its pole and $\MSbar$ masses,
the Bjorken sum rules for polarized electroproduction
($\Idx [g_1^{\rm ep}(x, Q^2) - g_1^{\rm en}(x, Q^2)]$)
\cite{Bjorken:1966jh}--\cite{Larin:1991tj},
and deeply inelastic neutrino-nucleon scattering
($\Idx[F_1^{\rm\overline{\nu}p}(x, Q^2)-F_1^{\rm\nu p}(x, Q^2)]$)
\cite{Bjorken:1967px}--\cite{Larin:1991zw},
and the static quark potential ($V(Q^2)$)
\cite{Susskind:1976pi}-\cite{Buchmuller:1980bm} and
\cite{Peter:1997ig}-\cite{Schroder:1999vy}.

We note that the non-singlet part of the Ellis-Jaffe sum rule
\cite{Ellis:1974kp}--\cite{Larin:1997qq} and \cite{Larin:1991tj}
gives the same scale as the former Bjorken sum rule,
the Gross-Llewellyn Smith sum rule 
\cite{Gross:1969jf,Bardeen:1978yd,Gorishnii:1986xm,Larin:1991tj}
the same scale as the latter.

All but the first scales are from known $\MSbar$ series, and for these 
at least the fermion vacuum polarization graphs must be given 
to two loops.  In several cases higher logarithmic moments are known, 
which will allow us to test the consistency of the procedure.  
We find four cases where the second-order formula gives the preferred scale. 

When the first-order solution is appropriate, the second moment gives a 
rough measure of the range in momenta which flow through the gluon, with
\beq
\label{delq}
  {\Delta q\over q} \approx 
  \half\left(\Delta_2\over \vev{f}\right)^\half \, .
\eeq
Here, $\Delta q$ is the standard deviation in $q$
and $\Delta_2$ is defined in \eq{deldef}.
A large range results in large coefficients at higher orders, as discussed
above.  When the second-order scale is appropriate, \eq{delq} clearly is not.
However, if higher moments are available, we may estimate this range using
\beq
\label{delqodd}
{\Delta q\over q} \approx 
   \left| {\Delta_{n+2}\over 4(n+1)\Delta_n} \right|^\half\, ,
\eeq
with $n$ odd.  In \Table{aptable}, we use $n=1$.  This expression
gives the standard deviation in a distribution modelled by a Gaussian 
times $\left[\logqs - \logq\right]^n$ to render it odd.  We found it to give 
reasonably consistent results for various $n$ when applied to several 
examples discussed below, though other measures are certainly possible.

For $R_{e^+ e^-}$, $M/\overline{M}$, and both
$\Gamma(B \rightarrow X_u e \overline{\nu})$ and $\Gamma(t \rightarrow b W )$
expressed in terms of $\MSbar$ masses, we find that the second-order
scale is appropriate, leading to significant corrections to the
anomalously low first-order scales, especially in the latter three.  
While the new scale for $M/\overline{M}$ is significantly increased, we note 
that $\Delta q/q$ is still relatively large, indicating sensitivity to
low-momentum scales even when $M$ is large, and threatening a 
poorly behaved series.  This apparently
infects the $b$ and $t$ decay rates when expressed in terms of pole 
masses, as shown by their low scales.  By contrast, $\MSbar$ masses behave 
more as bare masses, being sensitive to short distances; expressing the 
two decays in terms of these significantly improves their 
behavior~\cite{Smith:1995ev,Bigi:1994em,Smith:1994id,Beneke:1995qe,Ball:1995ni}.
This is clear from both their scales and widths.  
Both these series should be well-behaved, and well-represented by $\alphas$
at a single, physically reasonable scale.  But it is necessary to use 
second-order scale setting to see this; the first-order $q^*$
for each indicates a scale which is misleadingly low.

\begin{figure}
\epsfig{file=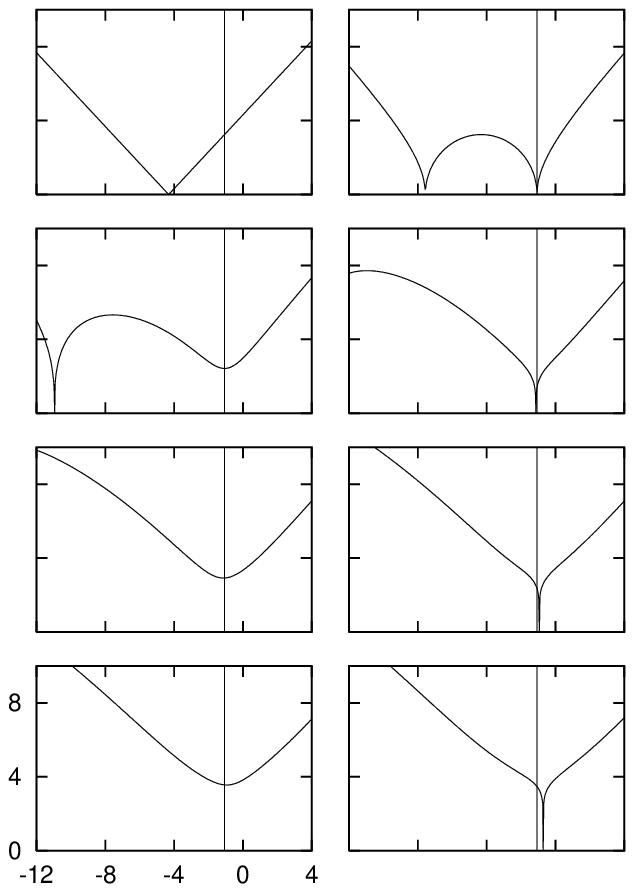, height=4.0in, width=4.0in}
\caption{The moments $\left|\Vev{\left[\logqsms - \logq\right]^n}\right|^{1/n}$
         as functions of $\logqsmsMb$  for 
         $n=1$ to $8$ (left to right) for $\Gamma(B\rightarrow X_u e 
         \overline{\nu}_e)$
         over the $\MSbar$ mass $\Mbmsbar$.  
         The vertical line indicates the choice for the scale $\logqsmsMb$
         using \eq{qstarext}.
         By choosing the second-order prescription such that the second moment
         vanishes, $\logqsmsMb$ is either near the minimum or the zero for all
         higher moments, minimizing higher-order terms in \eq{anseriesC}.}
\label{Bmoments}
\end{figure}

\Refs{Beneke:1995qe,Ball:1995ni,Ball:1995wa} provide very useful
values for fermion vacuum polarization contributions, and therefore 
logarithmic moments, computed to eighth order for the pole
to $\MSbar$ ratio, $\tau$, $B$ and $t$ decays.  These allow
us to compute their $\Delta q$'s using \eq{delqodd}, but more importantly,
to confirm the general picture as discussed in \sect{highord}.
In \fig{Bmoments} we use this information to display the first eight 
moments $\left|\Vev{\left[\logqsms - \logq\right]^n}\right|^{1/n}$ as 
functions of $\logqsmsMb$ for 
\mbox{$\Gamma(B\rightarrow X_u e \overline{\nu}_e)$} 
expressed in terms of the $\MSbar$ mass $\Mbmsbar$.  Here $q^{*2}_\msbar$
absorbs the fermion loop constant associated with the $\MSbar$ prescription,
as in \eq{eq:msVdiff}, and $\Mb$ is the $b$-quark pole mass.  We observe that 
choosing $\logqsmsMb$ to set the second moment to zero using 
\eq{qstarext} not only removes it and minimizes the third moment, 
it also sets all of the higher moments near their minima or zeros.  
It is clear that this is the natural scale for this process, and
that terms beyond second order which dress the leading gluon should 
be small.  The first moment
is clearly anomalous, and setting it to zero using \eq{firstorder} 
would evidently lead to large higher-order corrections.  
In general, we expect $f(q)$ to be either roughly even 
or odd about its typical scale $q^*$, and the sign of the second
moment, $\sigma^2$, should distinguish the two.  
For $\Delta q$ sufficiently small, using \eq{firstorder} or \eq{qstarext} 
depending on the sign of $\sigma^2$ should give reasonable values
except in very rare cases.

\begin{figure}
\begin{tabular}{cc}
\epsfig{file=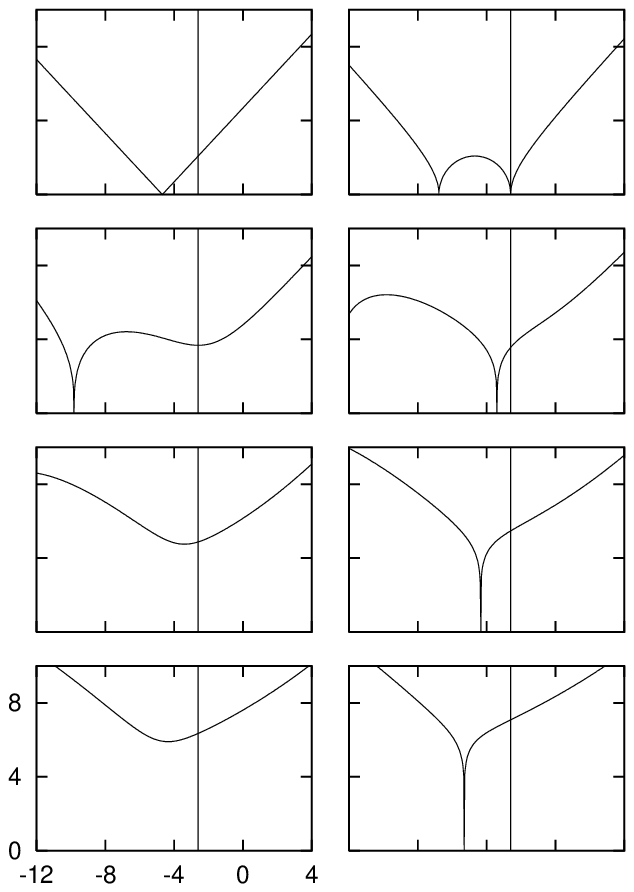, height=4.0in, width=4.0in}
\end{tabular}
\caption{The moments $\left|\Vev{\left[\logqsms - \logq\right]^n}\right|^{1/n}$
         as functions of $\logqsmsM$  for 
         $n=1$ to $8$ (left to right) for $M/\overline{M}$.
         The vertical line indicates the choice for the scale $\logqsmsM$
         using \eq{qstarext}.
        }
\label{Mmoments}
\end{figure}

The picture for $M/\overline{M}$, \fig{Mmoments}, is similar.
While choosing the second order scale is more appropriate than first,
causing the second moment to vanish and minimizing the third, the
zeros and minima of higher moments drift progressively lower.
Such behavior is anticipated by the relatively large value of
$\Delta q/q$, which indicates a wide range of contributing momenta.
In this case, higher moments are increasingly sensitive to 
lower $q$, and the corresponding coefficients will progressively 
increase.  We might improve the convergence of the series by 
methods discussed in \sect{improving}.  For example, choosing $q^*$ 
separately at each odd order in $\alphas$, causing the following even 
moment to vanish and miminizing the subsequent odd moment, with each 
$q^*$ indicating the characteristic scale for that moment.  An alternative
is to resum the entire set of polarization 
diagrams~\cite{Beneke:1995qe,Neubert:1995vb}.
Regardless, the ability to detect sensitivity to a large range of momenta,
in addition to the scale itself, by computing the first few logarithmic
moments is sufficient to warn of large higher order corrections.  In
this case, it suggests using $\Mmsbar$ rather than $M$ in expressions
for $b$ and $t$ decays.

\begin{table}
\caption{
Applications of second-order scale setting to several
processes.  The coefficients $c_n$ are defined in \sect{knownsrs}.
$q^*_1$ gives the scale set by \eq{firstorder}. $q^*_2$ gives
the preferred scale by \eq{qstarext} where appropriate,
also indicated by boxes.  $\Delta q$ measures the range of momentum running 
through the gluon.
}
\begin{center}
\begin{tabular}{cccccc}

\hline 
$c_1/c_0$ & $q^*_1$ & $c_2/c_0$ & $\sigma^2$ & $q^*_2$ & $\Delta q$ \\
\hline 

\\[0.0cm]
\multicolumn{6}{l}{
\underline{
$-\log W_{11}$:}}\\[.2cm]
$2.448$ & $3.402/a$ & $6.316$ & $0.3194$ & -- & $0.96/a$ \\[-.1cm]

\\[0.0cm]
\multicolumn{6}{l}{
\underline{
$R_{e^+ e^-}(s)$:}}\\[.2cm]
$-.691772$ & $0.7076 \sqrt{s}$ & $-.186421$ & $-.66497$ & 
\fbox{$1.064 \sqrt{s}$} & -- \\[-.1cm]

\\[0.0cm]
\multicolumn{6}{l}{
\underline{
$M/\overline{M}$:}}\\[.2cm]
$-4.6862$ & $0.09603 M$ & $17.623$ & $-4.3374$ & \fbox{$0.27205 M$} & 
$0.38 M$ \\[-.1cm]

\\[0.0cm]
\multicolumn{6}{l}{
\underline{
$R_\tau$:}}\\[.2cm]
$-2.2751$ & $0.32060 \Mtau$ & $5.6848$ & $0.50872$ & -- & 
$0.11 \Mtau$  \\[-.1cm]

\\[0.0cm]
\multicolumn{6}{l}{
\underline{
$\Gamma(B \rightarrow X_u e \overline{\nu})/\Mb^5$:}}\\[.2cm]
$-5.3382$ & $0.06932 \Mb$ & $34.410$ & $5.9139$ & -- & $0.084 \Mb$ \\[-.1cm]

\\[0.0cm]
\multicolumn{6}{l}{
\underline{
$\Gamma(B \rightarrow X_u e \overline{\nu})/\Mbmsbar^5$:}}\\[.2cm]
$-4.3163$ & $0.11554 \Mb$ & $8.0992$ & $-10.531$ & \fbox{$0.58534 \Mb$} & 
$0.35 \Mb$ \\[-.1cm]

\\[0.0cm]
\multicolumn{6}{l}{
\underline{
$\Gamma(t \rightarrow b W )/\Mt^3$:}}\\[.2cm]
$-4.2054$ & $0.12213 \Mt$ & $23.046$ & $5.3611$ & -- & $0.14 \Mt$ \\[-.1cm]

\\[0.0cm]
\multicolumn{6}{l}{
\underline{
$\Gamma(t \rightarrow b W )/\Mtmsbar^3$:}}\\[.2cm]
$-5.7076$ & $0.05763 \Mt$ & $6.0996$ & $-26.477$ & \fbox{$0.75502 \Mt$} & 
$0.34 \Mt$ \\[-.1cm]

\\[0.0cm]
\multicolumn{6}{l}{
\underline{
$\Idx [g_1^{\rm ep}(x, Q^2) - g_1^{\rm en}(x, Q^2)]$:}}\\[.2cm]
$-2$ & $e^{-1} Q = 0.3679 Q$ & $115/18$ & $43/18$ & -- & $0.28 Q$  \\[-.1cm]

\\[0.0cm]
\multicolumn{6}{l}{
\underline{
$\Idx[F_1^{\rm\overline{\nu}p}(x, Q^2)-F_1^{\rm\nu p}(x, Q^2)]$:}}\\[.2cm]
$-8/3$ & $e^{-4/3} Q = 0.2636 Q$ & $155/18$ & $3/2$ & -- & $0.16 Q$  \\[-.1cm]

\\[0.0cm]
\multicolumn{6}{l}{
\underline{
$V(Q^2)$:}}\\[.2cm]
$-5/3$ & $e^{-5/6} Q = 0.4346 Q$ & $25/9$ & $0$ & -- & $0$  \\[-.1cm]

\\[0.0cm]
\hline \hline 

\end{tabular}
\end{center}
\label{aptable}
\end{table}

\subsection{Quark mass and energy renormalization in lattice NRQCD}
\label{sec:nrqcd}

Lattice Nonrelativistic QCD (NRQCD) is an effective field theory designed to 
reproduce the results of continuum QCD for a heavy quark at energies 
small relative to its mass~\cite{Caswell:1986ui,Thacker:1991bm,Lepage:1992tx}.
Higher-dimensional operators provide systematic corrections ordered by quark 
velocity $v$ and lattice spacing $a$, and account for radiative processes 
above the cutoff, typically around the mass.  For a cutoff much larger than 
$\Lambda_{\rm QCD}$, lattice perturbation theory should give reliable values 
for the coefficients of these operators as well as the renormalization 
factors which connect bare to physical quantities.  \Ref{Lepage:1993xa} 
demonstrates that this expectation is valid, provided one uses a renormalized 
rather than bare coupling constant, and divides link gauge fields by their 
mean value to remove large tadpole contributions peculiar to the lattice.

\Refs{Davies:1992py} to \cite{Morningstar:1994qe}
present calculations of two of these quantities to first order in $\alphas$:
the renormalization factor $Z_m$, which connects the bare lattice heavy quark 
mass to its pole mass, and $E_0$, the shift from zero of the 
nonrelativistic energy of a heavy quark at rest.  
\Ref{Morningstar:1994qe}, using an action improved to \order{$v^2$} and 
\order{$a^2$}, and to \order{$v^4$} for spin-dependent interactions, found 
that first-order scale setting produced anomalous results for certain values 
of the bare mass, particularly after tadpole improvement.

\begin{figure}
\epsfig{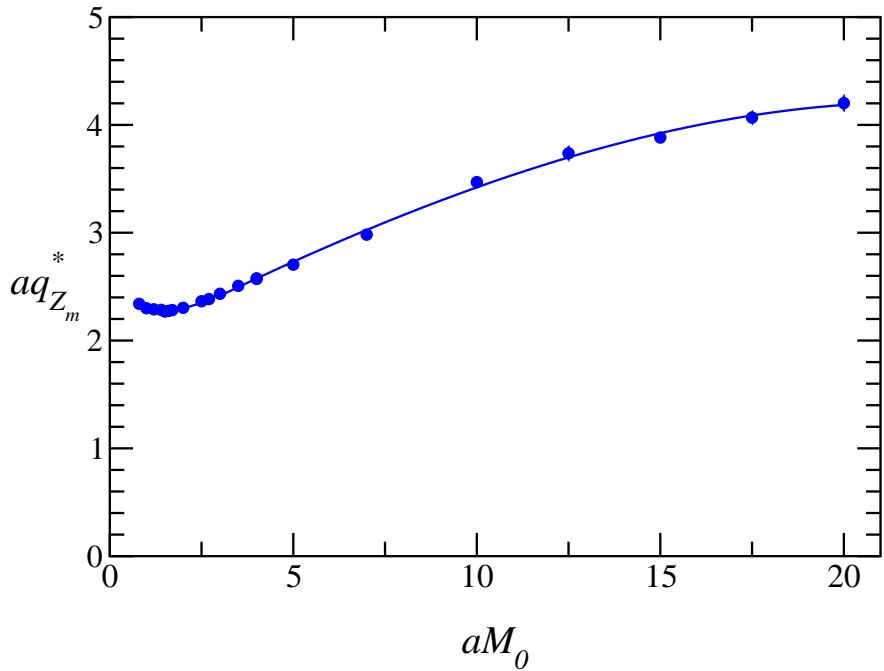}
\caption{
The BLM scale $q^*$ for the pole mass renormalization factor $Z_m$ as a 
function of the bare lattice mass $a M_0$ in NRQCD without tadpole 
improvement.  The first order solution determines $q^*$ for all values.
}
\label{nrmass}
\end{figure}

\begin{figure}
\epsfig{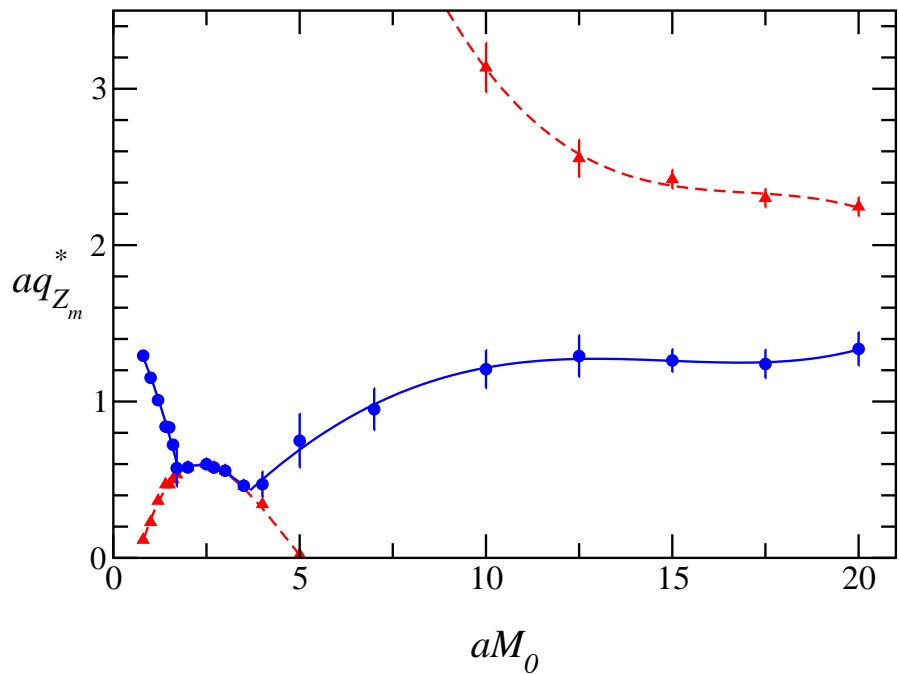}
\caption{
The BLM scale $q^*$ for the pole mass renormalization factor $Z_m$ as a 
function of the bare lattice mass $a M_0$ in NRQCD with tadpole improvement.  
The first order solution determines $q^*$ between $aM_0 = 2.00$ and $3.50$, 
the second order elsewhere.  Circles indicate the appropriate scale; triangles 
indicate the first-order solution in regions where it does not apply.  
}
\label{nrmassTI}
\end{figure}

\begin{figure}
\epsfig{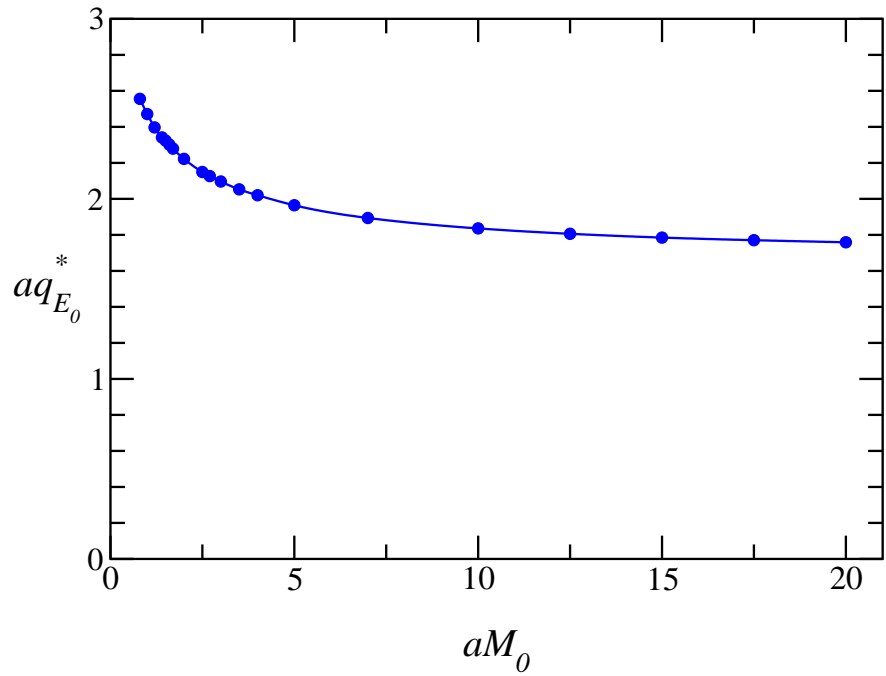}
\caption{
The BLM scale $q^*$ for the energy shift $E_0$ as a function of bare lattice
mass $a M_0$ in NRQCD without tadpole improvement.  The first order solution
determines $q^*$ for all values.
}
\label{nrenergy}
\end{figure}

\begin{figure}
\epsfig{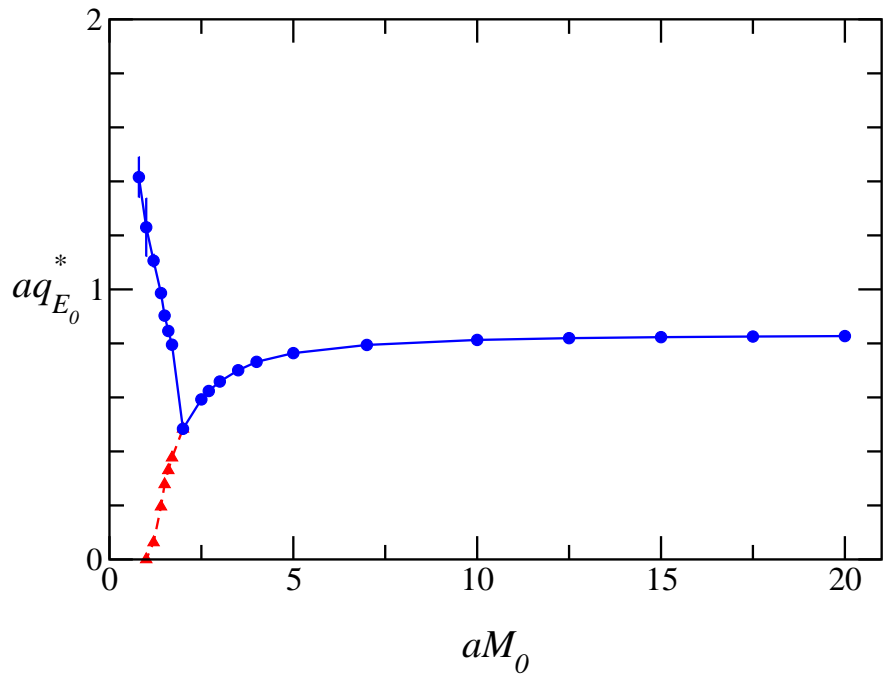}
\caption{
The BLM scale $q^*$ for the energy shift $E_0$ as a function of bare lattice
of the bare lattice mass $a M_0$ in NRQCD with tadpole improvement.  The first
order solution determines $q^*$ between $aM_0 = 0.80$ and $1.70$, the second 
order elsewhere.  Circles indicate the appropriate scale; triangles indicate 
the first-order solution in regions where it does not apply.  
}
\label{nrenergyTI}
\end{figure}

In Tables \ref{Mtable} to \ref{ETItable} and Figures \ref{nrmass} to 
\ref{nrenergyTI}, we present new values for the scale for a variety of bare 
quark masses $M_0$, both with and without tadpole improvement.  By applying 
\eq{qstarext} in regions where appropriate, we obtain a reasonable scale for 
all values of $M_0$, correcting the anomalies observed in 
\Ref{Morningstar:1994qe}.  As expected, there is a significant 
reduction in the scale after tadpole improvement.  The tadpole contributions 
to these renormalizations are quadratically divergent in the inverse lattice 
spacing, and so are generally large and sensitive to large momenta.  Tadpole 
improvement is designed to remove the bulk of these contributions, and so 
reduces the typical scale from 2 -- 4 to 0.5 -- 1.5 in units of the inverse 
lattice spacing $a$.

\begin{table}
\caption{
The BLM scale for the pole mass renormalization factor $Z_m$
for several values of the bare lattice mass $a M_0$ in NRQCD without tadpole 
improvement.  $a q^*_1$ gives the scale set by \eq{firstorder} in units of 
the inverse lattice spacing. $a q^*_2$ gives
the preferred scale by \eq{qstarext} where appropriate.
The parameter $n$ is set to ensure the stability of heavy quark propagator
evolution in simulations~\protect\cite{Thacker:1991bm}.
}
\begin{center}
\begin{tabular}{ccccccc}

\hline
 $n$ &  $aM_0$ &  $\vev{f}\equiv (Z_m - 1)/\alphas$  & $\vev{f\logaq}$ & 
$\vev{f\logaqtwo}$ & $aq^*_1$ & $aq^*_2$ \\
\hline
  1 & 20.00  &  0.4679(39) &  1.343(11)  &  4.091(62) &  4.202(71)  &  -- \\
    & 17.50  &  0.4860(34) &  1.364(10)  &  4.196(65) &  4.068(58)  &  -- \\
    & 15.00  &  0.5125(30) &  1.3907(85) &  4.211(41) &  3.883(44)  &  -- \\
    & 12.50  &  0.5410(46) &  1.426(14)  &  4.242(63) &  3.736(65)  &  -- \\
    & 10.00  &  0.5880(35) &  1.463(12)  &  4.384(68) &  3.469(43)  &  -- \\
    &  7.00  &  0.7057(17) &  1.5423(58) &  4.880(32) &  2.982(15)  &  -- \\
    &  5.00  &  0.8624(14) &  1.7153(48) &  5.538(29) &  2.7034(87) &  -- \\
    &  4.00  &  1.0071(27) &  1.9021(70) &  6.062(32) &  2.571(11)  &  -- \\
  2 &  4.00  &  1.0177(23) &  1.9264(80) &  6.164(33) &  2.577(11)  &  -- \\
    &  3.50  &  1.1268(23) &  2.0710(83) &  6.614(53) &  2.507(10)  &  -- \\
    &  3.00  &  1.2853(21) &  2.2859(68) &  7.351(35) &  2.4333(74) &  -- \\
    &  2.70  &  1.4119(19) &  2.4540(79) &  7.911(34) &  2.3846(73) &  -- \\
    &  2.50  &  1.5188(23) &  2.6141(74) &  8.390(52) &  2.3646(66) &  -- \\
    &  2.00  &  1.9018(22) &  3.1745(79) & 10.214(32) &  2.3039(53) &  -- \\
    &  1.70  &  2.2751(24) &  3.7546(75) & 12.039(56) &  2.2823(43) &  -- \\
    &  1.60  &  2.4384(24) &  4.0086(79) & 12.857(39) &  2.2750(41) &  -- \\
    &  1.50  &  2.6320(22) &  4.3141(78) & 13.745(37) &  2.2694(37) &  -- \\
  3 &  1.40  &  2.8804(23) &  4.7587(76) & 15.073(37) &  2.2842(34) &  -- \\
    &  1.20  &  3.5010(22) &  5.7986(83) & 18.143(39) &  2.2891(30) &  -- \\
    &  1.00  &  4.4915(19) &  7.4780(68) & 23.109(31) &  2.2990(19) &  -- \\
  5 &  0.80  &  6.3033(34) & 10.720(11)  & 32.131(63) &  2.3405(24) &  -- \\
\hline

\end{tabular}
\end{center}
\label{Mtable}
\end{table}

\begin{table}
\caption{
The BLM scale for the pole mass renormalization factor $Z_m$
for several values of the bare lattice mass $a M_0$ in NRQCD with tadpole 
improvement.  $a q^*_1$ gives the scale set by \eq{firstorder} in units of 
the inverse lattice spacing. $a q^*_2$ gives
the preferred scale by \eq{qstarext} where appropriate.
The parameter $n$ is set to ensure the stability of heavy quark propagator
evolution in simulations~\protect\cite{Thacker:1991bm}.
}
\begin{center}
\begin{tabular}{ccccccc}

\hline
 $n$ &  $aM_0$ &  $\vev{f}\equiv (Z_m - 1)/\alphas$  & $\vev{f\logaq}$ & 
$\vev{f\logaqtwo}$ & $aq^*_1$ & $aq^*_2$ \\
\hline
1 & 20.00 & -0.2381(39) & -0.385(11) & -0.367(62) &  2.246(61) & 1.34(11)  \\
  & 17.50 & -0.2224(34) & -0.371(10)  & -0.278(65) &  2.301(60) & 1.240(91) \\
  & 15.00 & -0.1996(30) & -0.3530(85) & -0.286(41) &  2.422(61) & 1.262(73) \\
  & 12.50 & -0.1773(46) & -0.333(14)  & -0.293(63) &  2.56(12)   & 1.29(13) \\
  & 10.00 & -0.1416(35) & -0.323(12)  & -0.223(68) &  3.14(16)   & 1.21(12) \\
  &  7.00 & -0.0566(17) & -0.3242(58) &  0.066(32) & 17.6(1.7)   & 0.95(13) \\
  &  5.00 &  0.0386(14) & -0.3018(48) &  0.335(29) &  0.0201(31) & 0.75(17) \\
  &  4.00 &  0.1126(27) & -0.2881(70) &  0.413(32) & 0.278(12)  & 0.650(59) \\
2 &  4.00 &  0.1232(23) & -0.2638(80) &  0.515(33) & 0.343(13)  & 0.471(81) \\
  &  3.50 &  0.1722(23) & -0.2664(83) &  0.586(53) &  0.461(12)  &    --    \\
  &  3.00 &  0.2381(21) & -0.2783(68) &  0.737(35) &  0.5574(85) &    --    \\
  &  2.70 &  0.2828(19) & -0.3107(79) &  0.781(34) &  0.5773(84) &    --    \\
  &  2.50 &  0.3180(23) & -0.3261(74) &  0.807(52) &  0.5988(74) &    --    \\
  &  2.00 &  0.4183(22) & -0.4581(79) &  0.846(32) &  0.5783(57) &    --    \\
  &  1.70 &  0.4899(24) & -0.6166(75) &  0.765(56) &  0.5329(44) & 0.57(12) \\
  &  1.60 &  0.5131(24) & -0.7058(79) &  0.699(39) & 0.5027(42) & 0.723(23) \\
  &  1.50 &  0.5376(22) & -0.8143(78) &  0.518(37) & 0.4689(37) & 0.835(17) \\
3 &  1.40 &  0.5795(23) & -0.8755(76) &  0.542(37) & 0.4698(34) & 0.839(15) \\
  &  1.20 &  0.6212(22) & -1.2529(83) & -0.043(39) & 0.3648(28) & 1.009(13) \\
  &  1.00 &  0.6518(19) & -1.9239(68) & -1.139(31) & 0.2286(16) & 1.152(11) \\
5 &  0.80 &  0.6964(34) & -3.009(11)  & -3.277(63) & 0.1153(15) & 1.293(24) \\
\hline

\end{tabular}
\end{center}
\label{MTItable}
\end{table}

\begin{table}
\caption{
The BLM scale for the energy shift $E_0$
for several values of the bare lattice mass $a M_0$ in NRQCD without tadpole 
improvement.  $a q^*_1$ gives the scale set by \eq{firstorder} in units of 
the inverse lattice spacing. $a q^*_2$ gives
the preferred scale by \eq{qstarext} where appropriate.
The parameter $n$ is set to ensure the stability of heavy quark propagator
evolution in simulations~\protect\cite{Thacker:1991bm}.
}
\begin{center}
\begin{tabular}{ccccccc}

\hline
 $n$ &  $aM_0$ &  $\vev{f}\equiv E_0/\alphas$  & $\vev{f\logaq}$ & 
$\vev{f\logaqtwo}$ & $aq^*_1$ & $aq^*_2$ \\
\hline
  1 & 20.00 & 2.25711(48) & 2.5490(13) & 10.8100(38) & 1.75884(55) & -- \\
    & 17.50 & 2.27657(44) & 2.6011(12) & 10.9496(40) & 1.77051(52) & -- \\
    & 15.00 & 2.30333(45) & 2.6694(13) & 11.1328(39) & 1.78507(54) & -- \\
    & 12.50 & 2.33892(69) & 2.7653(22) & 11.3882(87) & 1.80606(92) & -- \\
    & 10.00 & 2.39196(65) & 2.9062(23) & 11.7729(84) & 1.83583(92) & -- \\
    &  7.00 & 2.50419(41) & 3.1988(14) & 12.5623(49) & 1.89398(57) & -- \\
    &  5.00 & 2.64980(48) & 3.5784(15) & 13.5643(49) & 1.96446(59) & -- \\
    &  4.00 & 2.77360(93) & 3.8995(38) & 14.4035(99) & 2.0198(15)  & -- \\
  2 &  4.00 & 2.77159(78) & 3.8981(28) & 14.380(11)  & 2.0203(11)  & -- \\
    &  3.50 & 2.8585(10)  & 4.1118(34) & 14.952(11)  & 2.0529(13)  & -- \\
    &  3.00 & 2.97012(88) & 4.3973(34) & 15.670(12)  & 2.0965(13)  & -- \\
    &  2.70 & 3.0574(11)  & 4.6137(38) & 16.228(11)  & 2.1266(15)  & -- \\
    &  2.50 & 3.12485(90) & 4.7822(41) & 16.660(12)  & 2.1494(15)  & -- \\
    &  2.00 & 3.3492(12)  & 5.3483(43) & 18.116(15)  & 2.2221(16)  & -- \\
    &  1.70 & 3.5394(13)  & 5.8316(43) & 19.302(14)  & 2.2791(15)  & -- \\
    &  1.60 & 3.6163(13)  & 6.0299(45) & 19.843(16)  & 2.3018(16)  & -- \\
    &  1.50 & 3.7057(14)  & 6.2486(45) & 20.406(17)  & 2.3236(16)  & -- \\
  3 &  1.40 & 3.7865(16)  & 6.4417(61) & 20.873(18)  & 2.3411(21)  & -- \\
    &  1.20 & 4.0175(17)  & 7.0245(61) & 22.344(17)  & 2.3970(20)  & -- \\
    &  1.00 & 4.32658(96) & 7.8282(35) & 24.418(11)  & 2.4711(11)  & -- \\
  5 &  0.80 & 4.6581(20)  & 8.7418(74) & 26.784(27)  & 2.5557(23)  & -- \\
\hline

\end{tabular}
\end{center}
\label{Etable}
\end{table}

\begin{table}
\caption{
The BLM scale for the energy shift $E_0$
for several values of the bare lattice mass $a M_0$ in NRQCD with tadpole 
improvement.  $a q^*_1$ gives the scale set by \eq{firstorder} in units of 
the inverse lattice spacing. $a q^*_2$ gives
the preferred scale by \eq{qstarext} where appropriate.
The parameter $n$ is set to ensure the stability of heavy quark propagator
evolution in simulations~\protect\cite{Thacker:1991bm}.
}
\begin{center}
\begin{tabular}{ccccccc}

\hline
 $n$ &  $aM_0$ &  $\vev{f}\equiv E_0/\alphas$  & $\vev{f\logaq}$ & 
$\vev{f\logaqtwo}$ & $aq^*_1$ & $aq^*_2$ \\
\hline
1 & 20.00 &  1.05283(48) & -0.3998(13) &  3.2048(38) & 0.82706(52) &   --   \\ 
  & 17.50 &  1.04985(44) & -0.4027(12) &  3.2027(40) & 0.82549(50) &   --   \\
  & 15.00 &  1.04669(45) & -0.4076(13) &  3.1970(39) & 0.82306(51) &   --   \\
  & 12.50 &  1.04040(69) & -0.4143(22) &  3.1878(87) & 0.81948(89) &   --   \\
  & 10.00 &  1.03060(65) & -0.4272(23) &  3.1758(84) & 0.81281(90) &   --   \\
  &  7.00 &  1.00820(41) & -0.4643(14) &  3.1150(49) & 0.79431(56) &   --   \\
  &  5.00 &  0.97428(48) & -0.5243(15) &  2.9833(49) & 0.76409(58) &   --   \\
  &  4.00 &  0.94100(93) & -0.5878(38) &  2.8305(99) & 0.7318(15)  &   --   \\
2 &  4.00 &  0.93900(78) & -0.5891(28) &  2.807(11)  & 0.7307(11)  &   --   \\
  &  3.50 &  0.9137(10)  & -0.6502(34) &  2.671(11)  & 0.7006(13)  &   --   \\
  &  3.00 &  0.87572(88) & -0.7311(34) &  2.444(12)  & 0.6588(13)  &   --   \\
  &  2.70 &  0.8466(11)  & -0.7996(38) &  2.266(11)  & 0.6236(15)  &   --   \\
  &  2.50 &  0.82102(90) & -0.8590(41) &  2.111(12)  & 0.5927(15)  &   --   \\
  &  2.00 &  0.7312(12)  & -1.0621(43) &  1.583(15)  & 0.4837(16)  &   --   \\
 & 1.70 & 0.6442(13) & -1.2576(43) &  1.018(14)  & 0.3768(15)  & 0.7950(59) \\
 & 1.60 & 0.6056(13)  & -1.3421(45) &  0.830(16)  & 0.3302(14)  & 0.8460(65) \\
 & 1.50 & 0.5641(14)  & -1.4439(45) &  0.566(17)  & 0.2781(14)  & 0.9029(73) \\
3 & 1.40 & 0.4953(16) & -1.6171(61) &  0.088(18)  & 0.1955(16)  & 0.986(12)  \\
 & 1.20 & 0.3523(17)  & -1.9501(61) & -0.802(17)  & 0.06282(99) & 1.106(24)  \\
 & 1.00 & 0.13779(96) & -2.4285(35) & -2.035(11) & 0.0001489(93) & 1.23(11)  \\
5 & 0.80 & -0.3161(20) & -3.4380(74) & -4.629(27)  & 230.0(8.4)  & 1.416(75) \\
\hline

\end{tabular}
\end{center}
\label{ETItable}
\end{table}

\section{Conclusions}

In this paper we have derived a method which incorporates information
from higher orders into
the general prescription of \Ref{Brodsky:1983gc} for choosing 
the optimal scale $q^*$ for the strong coupling constant $\alphas$.  
We find that it corrects erroneous scales where the 
leading term or terms are anomalously small.  

The extended prescription states that \eq{qstarext} determines
the optimal scale $q^*$ when the argument of the square root
is positive.  When it is not, the first order formula in \eq{firstorder}
applies.  The choice of sign for the second-order solution should
be apparent either from continuity, or by checking that the solution
minimizes the next higher (cubic) moment in \eq{anseriesC} if it
is available.  In addition, higher moments give a measure of the 
range $\Delta q$ of momenta which flow through the gluon, and can confirm 
that the $q^*$ chosen in either case is indeed typical.  Large values
for the relative range $\Delta q/q$ can indicate large higher-order
contributions even when $q^*$ is large.

Our second-order prescription has several advantages.  It requires
a simple extension to the calculation, either numeric and analytic,
needed to implement the first-order prescription, requiring only
computation of an additional logarithmic moment.  Calculation of
higher moments can then help to further characterize the diagram
and confirm the scale choice.  It can also identify cases where
the first two terms are anomalously small, though such cases
are apparently rare.  It is appropriate regardless of
the number of loops included in the running coupling, and 
is not limited to the strong interactions.  Finally, it 
remedies erroneous scales in a variety of processes.

\section*{Acknowledgments}
We are grateful to S.~Brodsky, C.~Davies and J.~Shigemitsu 
for helpful conversations.  K.~H.\ thanks the members of the Newman Laboratory
theory group, Cornell, for their hospitality.  This work was supported in
part by grants from the National Science Foundation and the Department of 
Energy.

\end{document}